\input harvmac

\def\ZZ{\hbox{Z\kern-.4emZ}}
\def\RR{\hbox{R\kern-.6emR}}
\def\ZZs{\hbox{\zfonteight Z\kern-.4emZ}}

\def\tilde{\widetilde}

\lref\DixonIZ{
   L.~J.~Dixon and J.~A.~Harvey,
   ``String Theories In Ten-Dimensions Without Space-Time
Supersymmetry,''
   Nucl.\ Phys.\ B {\bf 274}, 93 (1986).
}

\lref\GiveonZZ{
 A.~Giveon and A.~Sever,
 ``Strings in a 2-d extremal black hole,''
 JHEP {\bf 0502}, 065 (2005)
 [arXiv:hep-th/0412294].
}

\lref\KutasovZM{
 D.~Kutasov, E.~J.~Martinec and M.~O'Loughlin,
 ``Vacua of M-theory and N=2 strings,''
 Nucl.\ Phys.\ B {\bf 477}, 675 (1996)
 [arXiv:hep-th/9603116].
}

\lref\KutasovVH{
 D.~Kutasov and E.~J.~Martinec,
 ``M-branes and N = 2 strings,''
 Class.\ Quant.\ Grav.\  {\bf 14}, 2483 (1997)
 [arXiv:hep-th/9612102].
}

\lref\OsorioHI{
  M.~A.~R.~Osorio and M.~A.~Vazquez-Mozo,
  Phys.\ Lett.\ B {\bf 280}, 21 (1992)
  [arXiv:hep-th/9201044].
}

\lref\OsorioVG{
  M.~A.~R.~Osorio and M.~A.~Vazquez-Mozo,
  Phys.\ Rev.\ D {\bf 47}, 3411 (1993)
  [arXiv:hep-th/9207002].
}

\lref\SeibergBY{
   N.~Seiberg and E.~Witten,
   ``Spin Structures In String Theory,''
   Nucl.\ Phys.\ B {\bf 276}, 272 (1986).
}

\lref\KutasovUA{
   D.~Kutasov and N.~Seiberg,
   ``Noncritical Superstrings,''
   Phys.\ Lett.\ B {\bf 251}, 67 (1990).
}

\lref\BouwknegtVA{ P.~Bouwknegt, J.~G.~McCarthy and K.~Pilch, `BRST
analysis of
physical states for 2-D (super)gravity coupled to (super)conformal
matter,''
arXiv:hep-th/9110031.
}

\lref\PandaGE{ S.~Panda and S.~Roy, ``BRST cohomology ring in $c(M) <
1$ NSR
string theory,'' Phys.\ Lett.\ B {\bf 358}, 229 (1995)
[arXiv:hep-th/9507054].
}

\lref\ItohIX{ K.~Itoh and N.~Ohta, ``Spectrum of two-dimensional
(super)gravity,'' Prog.\ Theor.\ Phys.\ Suppl.\  {\bf 110}, 97 (1992)
[arXiv:hep-th/9201034].
}

\lref\BouwknegtAM{ P.~Bouwknegt, J.~G.~McCarthy and K.~Pilch, ``Ground
ring for
the 2-D NSR string,'' Nucl.\ Phys.\ B {\bf 377}, 541 (1992)
[arXiv:hep-th/9112036].
}

\lref\ItohIY{ K.~Itoh and N.~Ohta, ``BRST cohomology and physical
states in 2-D
supergravity coupled to $c \le 1$ matter,'' Nucl.\ Phys.\ B {\bf 377},
113
(1992) [arXiv:hep-th/9110013].
}

\lref\ImbimboIA{
  C.~Imbimbo, S.~Mahapatra and S.~Mukhi,
  ``Construction of physical states of nontrivial ghost number in c < 1
string
  theory,''
  Nucl.\ Phys.\ B {\bf 375}, 399 (1992).
}

\lref\LianGK{ B.~H.~Lian and G.~J.~Zuckerman, ``New Selection Rules And
Physical States In 2-D Gravity: Conformal Gauge,'' Phys.\ Lett.\ B {\bf
254},
417 (1991).
}

\lref\WittenZD{
   E.~Witten,
   ``Ground ring of two-dimensional string theory,''
   Nucl.\ Phys.\ B {\bf 373}, 187 (1992)
   [arXiv:hep-th/9108004].
}

\lref\AdamsRB{
   A.~Adams, X.~Liu, J.~McGreevy, A.~Saltman and E.~Silverstein,
   ``Things fall apart: Topology change from winding tachyons'',
   arXiv:hep-th/0502021.
}

\lref\AtickSI{
   J.~J.~Atick and E.~Witten,
   ``The Hagedorn Transition And The Number Of Degrees Of Freedom Of
String
   Theory,''
   Nucl.\ Phys.\ B {\bf 310}, 291 (1988).
}

\lref\AlvarezUK{
   E.~Alvarez and M.~A.~R.~Osorio,
   ``Cosmological Constant Versus Free Energy For Heterotic Strings,''
   Nucl.\ Phys.\ B {\bf 304}, 327 (1988)
   [Erratum-ibid.\ B {\bf 309}, 220 (1988)].
}

\lref\DineVU{
  M.~Dine, P.~Y.~Huet and N.~Seiberg,
  ``Large And Small Radius In String Theory,''
  Nucl.\ Phys.\ B {\bf 322}, 301 (1989).
}

\lref\MaloneyRR{
   A.~Maloney, E.~Silverstein and A.~Strominger,
   ``De Sitter space in noncritical string theory,''
   [arXiv:hep-th/0205316].
}

\lref\McGreevyKB{
   J.~McGreevy and H.~Verlinde,
   ``Strings from tachyons: The c = 1 matrix reloaded,''
   JHEP {\bf 0312}, 054 (2003)
   [arXiv:hep-th/0304224].
}

\lref\KlebanovKM{
   I.~R.~Klebanov, J.~Maldacena and N.~Seiberg,
   ``D-brane decay in two-dimensional string theory,''
   JHEP {\bf 0307}, 045 (2003)
   [arXiv:hep-th/0305159].
}

\lref\TakayanagiSM{
   T.~Takayanagi and N.~Toumbas,
``A matrix model dual of type 0B string theory in two dimensions,''
   JHEP {\bf 0307}, 064 (2003)
   [arXiv:hep-th/0307083].
}

\lref\GukovYP{
   S.~Gukov, T.~Takayanagi and N.~Toumbas,
``Flux backgrounds in 2D string theory,''
   JHEP {\bf 0403}, 017 (2004)
   [arXiv:hep-th/0312208].
}

\lref\McGreevyDN{
J.~McGreevy, S.~Murthy and H.~Verlinde,
``Two-dimensional superstrings and the supersymmetric matrix model,''
   JHEP {\bf 0404}, 015 (2004)
   [arXiv:hep-th/0308105].
}

\lref\TakayanagiGE{
T.~Takayanagi,
``Comments on 2D type IIA string and matrix model,''
JHEP {\bf 0411}, 030 (2004)
[arXiv:hep-th/0408086].
}

\lref\GinspargBX{
   P.~H.~Ginsparg,
   ``Comment On Toroidal Compactification Of Heterotic Superstrings,''
   Phys.\ Rev.\ D {\bf 35}, 648 (1987).
}

\lref\AlvarezGaumeJB{
   L.~Alvarez-Gaume, P.~H.~Ginsparg, G.~W.~Moore and C.~Vafa,
   ``An O(16) X O(16) Heterotic String,''
   Phys.\ Lett.\ B {\bf 171}, 155 (1986).
}

\lref\McGuiganQP{
   M.~D.~McGuigan, C.~R.~Nappi and S.~A.~Yost,
   ``Charged black holes in two-dimensional string theory,''
   Nucl.\ Phys.\ B {\bf 375}, 421 (1992)
   [arXiv:hep-th/9111038].
}

\lref\HarveyNA{
   J.~A.~Harvey, D.~Kutasov and E.~J.~Martinec,
  ``On the relevance of tachyons,''
   [arXiv:hep-th/0003101].
}

\lref\LustTJ{
   D.~Lust and S.~Theisen,
   ``Lectures On String Theory,''
   Lect.\ Notes Phys.\  {\bf 346}, 1 (1989).
}

\lref\PolchinskiRR{
   J.~Polchinski,
   ``String theory. Vol. 2: Superstring theory and beyond,''
   Cambridge University Press (1998).
}

\lref\PolchinskiZF{
   J.~Polchinski,
   ``Evaluation Of The One Loop String Path Integral,''
   Commun.\ Math.\ Phys.\  {\bf 104}, 37 (1986).
}

\lref\GrossUB{
   D.~J.~Gross and I.~R.~Klebanov,
   ``One-Dimensional String Theory On A Circle,''
   Nucl.\ Phys.\ B {\bf 344}, 475 (1990).
}
\lref\DouglasUP{
   M.~R.~Douglas, I.~R.~Klebanov, D.~Kutasov, J.~Maldacena, E.~Martinec
and N.~Seiberg,
   ``A new hat for the c = 1 matrix model,''
   [arXiv:hep-th/0307195].
}

\lref\SeibergBX{
   N.~Seiberg,
   ``Observations on the moduli space of two dimensional string theory,''
   [arXiv:hep-th/0502156].
}

\lref\DixonAC{
   L.~J.~Dixon, P.~H.~Ginsparg and J.~A.~Harvey,
   ``(Central Charge C) = 1 Superconformal Field Theory,''
   Nucl.\ Phys.\ B {\bf 306}, 470 (1988).
}

\lref\AlvarezSJ{
   E.~Alvarez and M.~A.~R.~Osorio,
   ``Superstrings At Finite Temperature,''
   Phys.\ Rev.\ D {\bf 36}, 1175 (1987).
}

\lref\BrienPN{
   K.~H.~O'Brien and C.~I.~Tan,
   ``Modular Invariance Of Thermopartition Function And Global Phase
Structure
   Of Heterotic String,''
   Phys.\ Rev.\ D {\bf 36}, 1184 (1987).
}

\lref\SeibergEB{
  N.~Seiberg,
  ``Notes On Quantum Liouville Theory And Quantum Gravity,''
  Prog.\ Theor.\ Phys.\ Suppl.\  {\bf 102}, 319 (1990).
}

\lref\KlebanovQA{
   I.~R.~Klebanov,
   ``String theory in two-dimensions,''
   arXiv:hep-th/9108019.
}

\lref\GinspargIS{
   P.~H.~Ginsparg and G.~W.~Moore,
   ``Lectures on 2-D gravity and 2-D string theory,''
   arXiv:hep-th/9304011.
}

\lref\BergmanYP{
   O.~Bergman and S.~Hirano,
   ``The cap in the hat: Unoriented 2D strings and matrix(-vector)
models,''
   JHEP {\bf 0401}, 043 (2004)
   [arXiv:hep-th/0311068].
}

\lref\GomisVI{
   J.~Gomis and A.~Kapustin,
   ``Two-dimensional unoriented strings and matrix models,''
   JHEP {\bf 0406}, 002 (2004)
   [arXiv:hep-th/0310195].
}

\Title{
}
{\vbox{\centerline{Heterotic Strings in Two Dimensions}
\medskip
\centerline{and New Stringy Phase Transitions}}}
\medskip
\centerline{\it
Joshua L. Davis${}^{1}$, Finn Larsen${}^{1}$, and Nathan Seiberg${}^{2}$
}
\bigskip
\centerline{${}^1$Michigan Center for Theoretical Physics, Ann Arbor,
MI 48109}
\smallskip
\centerline{${}^2$School of Natural Sciences, Institute for Advanced
Study,
Einstein Drive, Princeton, NJ 08540}
\smallskip

\vglue .3cm
\bigskip\bigskip\bigskip
\centerline{\bf Abstract}

\noindent
  We discuss heterotic string theories in two dimensions with gauge
groups $Spin(24)$ and $Spin(8)\times E_8$. After compactification
the theories exhibit a rich spectrum of states with both winding
and momentum. At special points some of these stringy states
become massless, leading to new first order phase transitions. For
example, the thermal theories exhibit standard thermodynamics
below the phase transition, but novel and peculiar behavior above
it.  In particular, when the radius of the Euclidean circle is smaller than the
phase transition point the torus partition function is not given by the thermal
trace over the spacetime Hilbert space. The full moduli space of
compactified theories is $13$ dimensional, when Wilson lines are
included; the $Spin(24)$ and $Spin(8)\times E_8$ theories
correspond to distinct decompactification limits.

\Date{}

\newsec{Introduction and Summary}

There has recently been renewed interest in string theories with
two-dimensional target space \refs{\McGreevyKB,\KlebanovKM} (for
earlier work on string theory in two dimensions see {\it e.g.}
\refs{\KlebanovQA,\GinspargIS}). One aspect of this development is
that several new theories have been proposed
\refs{\TakayanagiSM\DouglasUP\GomisVI\BergmanYP\GukovYP\TakayanagiGE-
\SeibergBX}.

The goal of this paper is to study heterotic strings in two target
space dimensions. These are theories that couple to $(1,0)$
worldsheet supergravity. The supersymmetric side of the
world-sheet theory has the same structure as the $N=1$
non-critical superstring. The bosonic side matches the bosonic
fields of the supersymmetric side and has, in addition, a $c_L=12$
matter sector. This matter can be organized into either $Spin(24)$
or $Spin(8)\times E_8$ current algebras, thus defining two
consistent heterotic string theories. These theories were
discussed in \refs{\McGuiganQP,\GiveonZZ}. Other theories with
somewhat similar features were studied in
\refs{\OsorioHI\OsorioVG\KutasovZM-\KutasovVH}.

The spectrum of the $Spin(24)$ theory are $24$ massless
``tachyon'' fields, as well as discrete states. The propagating
modes of  the $Spin(8)\times E_8$ theory are $8_C$ massless
fermions of one chirality, $8_S$ massless fermions of the other
chirality, and $8_V$  massless ``tachyons.''

It is interesting to compactify the heterotic strings, with or without
twisting by its discrete symmetries. Then each field theory degree of
freedom gives rise to a tower of excitations {\it a la} Kaluza-Klein.
An important novelty is that, unlike the bosonic, type 0 and type II
string theories, we find infinitely many states which have both momentum
and winding. Thus there is a rich spectrum of ``states'' in the theory
with compact time, with each level transforming as an increasingly complex
representation of the gauge group. These modes can lead to interesting
phenomena.

Some of our theories exhibit self-duality under inversion of the
compactification radius $R$. At the self-dual points there are
enhanced gauge symmetries such that T-duality is part of the gauge
symmetry \DineVU. Due to the enhanced symmetry, there can be new
massless particles which can give rise to phase transitions.

The most striking effect occurs when some of the string theory
modes become massless (in the sense of one-dimensional Liouville
theory \SeibergEB). In these cases the torus amplitude is
non-analytic and the theory undergoes a phase transition. The mode
that becomes massless can be either a complex boson $\Phi$ or a
complex fermion $\chi$. We can describe its one-dimensional
Landau-Ginzburg mean field theory Lagrangian as
\eqn\lglag{\eqalign{ &\CL_\Phi = \half |\partial_\phi \Phi|^2 +
\half m(R)^2 \Phi^2 \cr &\CL_\chi =
i\chi^\dagger\partial_\phi\chi+ m(R) \chi^\dagger \chi
  }}
In our examples the mass $m(R)$ has a simple zero; specifically
$m(R)=\half(R-{1\over R})$. The one loop fluctuations of $\Phi$
and $\chi$ lead to finite, nonanalytic terms
  \eqn\lgonel{\eqalign{
&Z_\Phi = -\int {V_L dp \over 2\pi} \log(p^2 + m(R)^2) =  - V_L
|m(R)| + {\rm const}\cr &Z_\chi = \half \int {V_L dp \over 2\pi}
\log(p^2 + m(R)^2) = + \half V_L |m(R)|+ {\rm const}
  }}
Here $V_L=\int d\phi ~1$ is the size of the spatial direction
$\phi$.  The (infinite) constants are independent of
$m(R)$ and can be ignored. All our torus amplitudes are analytic
functions of $R$ plus possible terms arising from \lgonel. Our
results are summarized in table 2 and 3 in section 4.

The interpretation of the results raises conceptual issues that
are not fully understood; they are discussed in section 4.4.  For
example, we will see that the torus amplitude of a theory
compactified on a small thermal circle is not given by the
standard thermodynamical trace over the spacetime Hilbert space.
It is not clear whether there exists an alternate thermodynamical
description of the physics with such small radius.  The moduli
space of other compactifications might have a boundary at finite
radius beyond which the radius cannot be reduced.

We would like to clarify a few general points of potential
confusion.  The target space of the theories we consider have a
linear dilaton along the spatial direction, $\phi$, so that the
string coupling constant varies as
  \eqn\stringcou{g_s(\phi) = e^\phi}
We focus on the weakly coupled region of the target space where
the string coupling is arbitrarily small $g_s(\phi)\to 0$, while
the string scale $M_s$ is finite. There, the infrared dynamics of
the gauge theory can be ignored, because it is important only at
energies below $g_s(\phi) M_s\to 0$. The typical energy scale we
consider, including the scale $1/R$ set by our compactifications,
is of order string scale $M_s$ and, therefore, not affected by the
infrared dynamics.

We will be interested in the string theory partition function written
in the form
\eqn\stringpart{\int d\phi Z(\phi)= \int d\phi \left(
  e^{-2\phi} A_0 + A_1 +e^{2\phi} A_2 + ... \right)}
where the $\phi$ dependence is associated with the powers of the
string coupling \stringcou\ and, therefore, the coefficient $ A_n$
is the genus $n$ contribution.  The sphere term $A_0 $ is
proportional to the compactification radius $R$ and is not
interesting for our purpose. The torus amplitude $A_1$ is more
interesting; for example, it receives the non-analytic
contributions \lgonel. Importantly, $A_0$ and $A_1$ depend only on
physics in the weak coupling region, and on the weak coupling
spectrum. $A_n$ with larger $n$ depend on the details of the
interactions in the strong coupling region; but they are
negligible for $\phi \to - \infty$. Usually, one turns on a
tachyon background with coefficient $\mu$ to control the
perturbative expansion \stringpart\ but this will not be needed
here.

The theories we consider have discrete states formed from the
gauge currents; but the ground ring and its associated towers of
currents seems absent. In the bosonic and supersymmetric theories
such currents are related to the symmetries of the dual matrix
model description, specifically the symmetries expressing
incompressibility of the free fermion representation. (Some
discussions of discrete states and the ground ring in bosonic and
superstring theories are
\refs{\WittenZD\BouwknegtAM\BouwknegtVA\ItohIX\ItohIY\PandaGE\ImbimboIA-
\LianGK}.) The absence of this structure for the heterotic strings
indicates that, if a dual matrix model description exists at all,
it must have some significant new feature. Additionally, heterotic
strings support no D-brane boundary states. Since the modern
interpretation of matrix models identifies the matrix eigenvalues
with D-brane coordinates \refs{\McGreevyKB,\KlebanovKM}, this is
another indication that a matrix model description cannot be
simple. It would clearly be interesting to find a non-perturbative
formulation of the heterotic strings discussed here.

An illuminating way to explore heterotic theories is to employ lattice
technology.
In two dimensions the uncompactified theories can be classified by even
self-dual
lattices in $16$ Euclidean dimensions, using the covariant lattice
construction
(which includes right moving fermions and superconformal ghosts). This
confirms
that there really are exactly two ``fundamental" theories, with gauge
groups
$Spin(24)$ and $Spin(8)\times E_8$, respectively. This contrasts with
ten dimensions
where, in addition to the familiar supersymmetric $Spin(32)/\ZZ_2$ and
$E_8\times E_8$ theories,
there are a number of non-supersymmetric theories.

The lattice construction also shows that, after compactification, all
theories are
connected: there is a $13$ dimensional moduli space, parametrized by
the radius
of compactification and $12$ independent Wilson lines. This is much
richer
than for other strings in two dimensions. As illustrations, we show
explicitly how
the twisted lines of theories can be reinterpreted in terms of Wilson
lines; and how
T-duality relates the $Spin(24)$ and the $Spin(8)\times E_8$ theories,
after the
introduction of suitable Wilson lines.

The remainder of the paper is organized as follows. In section 2 we
define the basic
$Spin(24)$ and $Spin(8)\times E_8$ theories. We also discuss the
discrete symmetries
of the theories. In section 3 we consider compactifications of the two
theories, with or
without twisting of their discrete symmetries. T-duality and enhanced
symmetry points
are discussed as well. In section 4 we evaluate the torus
partition function explicitly for the different lines of theories and
discuss the phase
transition in detail. Finally, we include an appendix where lattice
constructions
are used to classify the theories and reconsider their
interconnections. Throughout
the paper we use units in which $\alpha^\prime =2$.

\newsec{Theories in Noncompact Space}

The right movers are the $\hat c=1$ noncritical string: a
Liouville field $\phi$, (Euclidean) time $x$, and their fermionic
superpartners $\psi_\phi$ and $\psi_x$. The slope of the Liouville
field is $Q=1$
which is such that it contributes $c_\phi=13$ to the
central charge. The left movers constitute
a noncritical bosonic theory that includes the Liouville
field, (Euclidean) time and, in order to have total
left moving central charge $26$, a $c_L=12$
bosonic CFT which we will take to be $24$ free fermions
$\bar\lambda^I$ with $I=1,...,24$. In the remainder
of this section we discuss the two natural theories constructed
out of these building blocks.

\subsec{$Spin(24)$ theory}

Here we correlate the spin structure of the free $24$ fermions
with that of the right movers. The physical vertex operators are
  \eqn\twefourv{\eqalign{
  &G={\cal J} \bar {\cal J}\cr
  &A^{IJ}=  {\cal J}\bar \lambda^I \bar\lambda^J\cr
  &T^I(k)=e^{-\varphi}\bar \lambda^I V_k
   }}
  where the operators
  \eqn\uone{\eqalign{
   &{\cal J}=e^{-\varphi}\psi_x\cr
  &\bar{\cal  J}=\bar \partial \bar x
}}
are $U(1)$ currents and the wave functions are
  \eqn\vldef{
  V_k = e^{i k(x+\bar x) +
  (1-|k|)(\phi+\bar \phi)}
} The absolute value in the coefficient of $\phi$ was explained in
\SeibergEB. The discrete states $G$ and $A^{IJ}$ are the two
dimensional graviton/dilaton and the $Spin(24)$ gauge fields.
$T^I(k)$ represent $24$ massless scalars ``tachyons.'' The Ramond
sector does not lead to physical particles because the $Spin(24)$
spin fields, $\bar S^\alpha$ and $\bar S^{\dot \alpha}$, have
dimension $\bar{\Delta}={3\over 2}$ and $V_k$ has dimensions
$(\Delta,\bar{\Delta})=(\half,\half)$ for all $k$. These fields,
however, will play a role when we discuss the compactified theory.
Clearly, the operators in \twefourv\ are mutually local. The
partition function (in the notation of \PolchinskiRR)
\eqn\fewterms{ Z_F (\bar{\tau}) = \half \left[ Z_0^0
(\bar{\tau})^{12}  - Z_1^0 (\bar{\tau})^{12} -
Z^1_0(\bar{\tau})^{12} \right] } is modular invariant. Also note
that $Z_F (\bar{\tau}) = 24$ is independent of $\bar \tau$
\GiveonZZ, and hence it is manifestly modular invariant.

It is possible to turn on background tachyons which break the
continuous symmetry $Spin(24)\to Spin(23)$.

\subsec{$Spin(8)\times E_8$ theory}

Here we divide the $24$ fermions into two groups: $\bar{\lambda}^i$ with
$i=1,...,8$ and $16$ other fermions. The latter lead to an $E_8$
left moving CFT.  The spin structure of the $\bar{\lambda}^i$ is
correlated with that of the right movers. In this theory the physical
vertex
operators are
  \eqn\soeightv{\eqalign{
  &G={\cal J} \bar {\cal J}\cr
  &A^{ij}=  {\cal J}\bar \lambda^i \bar\lambda^j\cr
  &A^{ab}={\cal J}\bar J^{ab}\cr
  &T^i(k)=e^{-\varphi}\bar \lambda^i V_k \cr
  &\Psi^\alpha=e^{-\half\varphi+i\half H}\bar S^\alpha V_k~~,\quad k\geq
0 \cr
  &\tilde\Psi^{\dot\alpha}=e^{-\half\varphi-i\half H}\bar
  S^{\dot\alpha} V_k ~~,\quad k\leq 0
  }}
Again, $\cal J$ and $\bar{\cal J}$ are the $U(1)$ currents \uone\
which lead to discrete states: $G$ is the graviton/dilaton,
$A^{ij} $ are the $Spin(8)$ gauge fields and $A^{ab}$ are the
$E_8$ gauge fields constructed from the $E_8$ currents $\bar
J^{ab}$.  The other vertex operators represent propagating
particles: $T^i$ is a scalar in $8_V$ of $Spin(8)$, $\Psi^\alpha$
is a left moving spacetime fermion in $8_S$, and $\tilde
\Psi^{\dot \alpha}$ is a right moving spacetime fermion in $8_C$.
Unlike the $Spin(24)$ theory, here, the dimension of the $Spin(8)$
fields $S^\alpha$ and $S^{\dot \alpha}$ is $\Delta = {1\over 2}$,
thus giving rise to physical fermions. The conditions on the
momentum $k$ arise from locality with respect to the world-sheet
supercurrent ({\it i.e.} the Dirac equation). Note that the
spectrum is anomaly free even though it is chiral. It is
straightforward to check that the operators in \soeightv\ are
mutually local. The partition function of the theory
\eqn\jacid{\eqalign{ Z_F(\bar{\tau}) &= \half \left[ Z_0^0
(\bar{\tau})^{4}  - Z_1^0 (\bar{\tau})^{4} - Z_0^1
(\bar{\tau})^{4} \right] \cdot\left( Z_0^0 (\bar{\tau})^8 + Z_0^1
(\bar{\tau})^8 + Z_1^0 (\bar{\tau})^8 \right)\cr &= [(8-8) +
(64-64)q+\cdots ]\cdot [1 + 248q+\cdots]=0 }} is modular
invariant. Note that in this case $Z_F\equiv 0$ identically (this
follows from the Jacobi identity, familiar from spacetime
supersymmetry in $10$ dimensions).

It is possible to turn on background tachyons which break the
continuous symmetry $Spin(8) \to Spin(7)$. The effective
Lagrangian can include a coupling of the form $T^i \Psi^\alpha
\tilde \Psi^{\dot \alpha} \gamma_{i\alpha\dot\alpha}$ with
possible derivatives. It is amusing that this spectrum of
particles is the same as in the worldsheet light cone description
of the IIA critical string. However, unlike that theory, here,
because of the linear dilaton, there is no two-dimensional
Lorentz invariance and no $(8,8)$ two dimensional supersymmetry.

\subsec{Discrete Symmetries}
We next discuss the discrete symmetries of the theories. We
focus on transformations that do not break the gauge symmetry.

The spacetime fermion number $F_R$ and the right moving
world-sheet fermion number $f_R$ are represented in the
same way as in the superstring \SeibergBX
\eqn\symmR{\eqalign{
& (-)^{F_R}: ~~~\varphi\to \varphi + 2\pi i~~\cr
& (-)^{f_R}:~~~\varphi\to \varphi +\pi i~,~~H \to H+\pi
}}
In the left moving sector we must proceed differently. The center of
$Spin(4n)$ is
$\ZZ_2\times \ZZ_2$. The generators of the center transform
representations according
to their $Spin(4n)$ conjugacy class
\eqn\spincn{\eqalign{
{\cal Z}_1 =(-)^{F_L}&: ~~{\cal O}_V\to {\cal O}_V~~~~~;~ {\cal O}_S\to
-{\cal O}_S~~~;~{\cal O}_C\to -{\cal O}_C~\cr
{\cal Z}_2 =(-)^{f_L} &: ~~{\cal O}_V\to -{\cal O}_V~~~;~ {\cal O}_S\to
{\cal O}_S~~~~~;~{\cal O}_C\to - {\cal O}_C
}}
while the ${\cal O}_0$ are invariant. The transformation ${\cal Z}_1$
is a
rotation by
$2\pi$ around some axis in the internal space. It is therefore natural
to define the
left moving spacetime fermion number as $(-)^{F_L}\equiv {\cal Z}_1$.
The transformation
${\cal Z}_2$ is world-sheet fermion number insofar as the $Spin(2n)$
current algebra is realized
in terms of $2n$ free fermions. It is therefore natural to define the
left-moving world-sheet
fermion number as $(-)^{f_L}\equiv {\cal Z}_2$. Note that this latter
identification also makes
sense for $Spin(8)\times E_8$, because there are always an even number
of $E_8$ fermions.

With these notations and conventions, the theories we consider are
defined with diagonal
GSO projections, {\it i.e.} the operators satisfy $(-)^{F_L+F_R}
=(-)^{f_L+f_R}=1$ on physical
states. In contrast, the elements $(-)^{F_L}$ and $(-)^{f_L}$ in the
center of $Spin(4n)$
act as symmetries. In view of the GSO projection the symmetries can
equally be characterized
in terms of right moving quantities $(-)^{F_R}$ or $(-)^{f_R}$.

There are two more discrete transformations of interest. Spacetime
parity ${\cal P}$ acts
as
\eqn\Pdef{\eqalign{
{\cal P}: ~~~H\to -H~;~~x\to -x~;~~{\bar x}\to -{\bar x}
}}
while charge conjugation ${\cal C}$ acts on the $Spin(4n)$ lattice as
\eqn\Cdef{\eqalign{
{\cal C}:~~~S\leftrightarrow C
}}
Neither of these transformations are symmetries of the $Spin(8)\times
E_8$ theory.
For example, ${\cal P}$ would transform $\Psi^\alpha$ in \soeightv\
into a state
with $k\leq 0$ and $H\to -H$, leaving ${\bar S}^\alpha$ intact; but
there is no
such state in the spectrum. However, the combined transformation ${\cal
C}{\cal P}$ {\it is}
a symmetry of the theory: it simply interchanges the spinorial vertex
operators $\Psi^\alpha$ and
$\tilde{\Psi}^{\dot\alpha}$. The ${\cal P}$, ${\cal C}$ and ${\cal
C}{\cal P}$ are all symmetries
of the $Spin(24)$ theory.

The diagonal element in the center of $Spin(4n)$, generated by
${\cal Z}_1{\cal Z}_2=(-)^{F_L+f_L}$, is related to ${\cal Z}_2$
through conjugation
by ${\cal C}{\cal P}$: we have ${\cal CP}(-)^{f_L} {\cal
CP}=(-)^{F_L+f_L}$ when acting
on any operator in the theories. Thus, $(-)^{F_L+f_L}$ acts in the same
way as $(-)^{f_L}$,
up to a change of conventions; so, later, it will be sufficient to
consider
orbifolds and twists by $(-)^{F_L}$ and $(-)^{f_L}$.

It is significant that the discrete symmetries are in fact elements of
the center of the group,
which can be continuously related to the identity. This means
compactifications twisted
by each of these symmetries are all connected to untwisted
compactification. We will
make this more explicit in the appendix.

As a final comment on discrete symmetries, recall that $Spin(8)$ allows
triality
transformations, realized as outer automorphisms of the algebra. One
element
of the triality group acts on the weight lattice by cyclic permutation
of the conjugacy
classes $V\to S\to C\to V$. Concretely, this means we can replace the
operators
appearing in \soeightv\ according to
\eqn\triality{
{\bar\lambda}^i\to {\bar S}^\alpha \to {\bar S}^{\dot\alpha} \to
{\bar\lambda}^i
}
It is important that no physical observable will be different in
theories related by
triality, because an automorphism just amounts to renaming of the
representations.

\subsec{Orbifolds}
Starting from the $Spin(24)$ and $Spin(8)\times E_8$ theories discussed
above
it is natural to seek new theories by orbifolding with respect to the
discrete
symmetries. In the following we argue that this does not lead to
interesting new
possibilities.

First, let us orbifold by $(-)^{F_L}$. As indicated in \spincn\  the
untwisted sector
is the $V$ conjugacy class of the $Spin(4n)$ and also the discrete
states, but
the R-sector (if any) is projected out. An R-sector arises from the
twisted states, but it
has the opposite correlation between spacetime chirality
and $Spin(4n)$ chirality. Explicitly, for the $Spin(8)\times E_8$
theory in \soeightv,
the $\Psi^\alpha$ and $\tilde{\Psi}^{\dot\alpha}$ are being replaced by
$\tilde{\Psi}^\alpha$ and ${\Psi}^{\dot\alpha}$. This does not lead to
a genuinely
new theory: it reduces to the transformation ${\cal C}$ introduced in
\Cdef.
In the $Spin(24)$ there are no R-states at all, so the orbifold leaves
the
theory invariant.

Next, let us consider orbifold by $(-)^{f_L}$. From \spincn\ we see
that the
untwisted sector consists of the $S$ conjugacy class along with the
discrete states
in the $0$ conjugacy class. In the $Spin(8)\times E_8$ theory this leaves the
propagating states
$\Psi^\alpha$ and so, adding the twisted states permitted by locality,
we find
the propagating states
\eqn\twfl{\eqalign{
&\Psi^\alpha=e^{-\half\varphi+i\half H}\bar S^\alpha V_k~~,\quad k\geq
0 \cr
&T^{\dot\alpha} =  e^{-\varphi} {\bar S}^{\dot\alpha}  V_k  \cr
&\tilde\Psi^i= e^{-\half\varphi-i\half H} \bar{\lambda}^i V_k~~,\quad
k\leq 0 \cr
}}
This is related to the original spectrum \soeightv\ by triality.

The $(-)^{f_L}$ orbifold of the $Spin(24)$ theory is formally expected
to
have the same structure as the $Spin(8)\times E_8$ theory but now,
because
the spin fields have dimensions $\bar{\Delta}=3/2$, the only
propagating states
are the $\tilde\Psi^i$, {\it i.e.} $24$ chiral fermions. This spectrum
is anomalous in
spacetime and the theory is probably inconsistent.

\newsec{Compactification}

In this section we discuss the compactification of the $Spin(24)$ and
$Spin(8)\times E_8$
theories on a circle of radius $R$. We also consider theories that are
twisted by discrete
symmetries that commute with the gauge symmetry. As discussed in
section 2.3, the
non-trivial twists are $(-)^{F_L}$ and $(-)^{f_L}$.

\subsec{Circle compactification: no twist}
The spectrum at generic radius $R$ includes the currents $\cal J$
and $\bar{\cal J}$ \uone\ and the discrete states $G$ and $A^{IJ} $ of
\twefourv\  as well as
\eqn\twer{\eqalign{
  &T_V=e^{-\varphi}\bar \CO_V(\half +nw)
V_{n,w}\cr
  &\Psi_S = e^{-\half\varphi + i\half H }\bar \CO_S(\half +nw) V_{n,w}~,
\qquad  p_R \ge 0  \cr
  &\tilde \Psi_C = e^{-\half\varphi - i\half H }\bar \CO_C(\half +nw)
V_{n,w}~, \qquad p_R  \le 0\cr
  }}
where now the wave function is
\eqn\wavef{
V_{n,w}=
e^{i {n\over R}(x+{\bar x}) + i {w\over 2} R(x-{\bar x}) +
(1-|p_R|)(\phi + {\bar \phi})}
}
with $p_{R} = {n\over R} + {wR\over 2}$. The $n$, $w$ are integers and
$\bar\CO_r(\bar \Delta)$ are operators in the conjugacy class $r=V, S,
C$ of
$Spin(4n)$ with dimension $\bar \Delta$. The spectrum \twer\ is modular
invariant
(some details of this are discussed in section 4.1).

The theories are clearly invariant under $R\to {2\over R}$.  The right
moving
$U(1)$ symmetry cannot be enhanced but, at the selfdual
radius $R=\sqrt 2$, the left moving $U(1)$ symmetry whose current is
$\bar
\partial \bar x$ is enhanced to $SU(2)$.

The list of operators in \twer\ represents schematically the spectrum
of either the
$Spin(24)$ theory or the $Spin(8)\times E_8$ theory. The difference
between the
theories appears when constructing the operators $\bar\CO_r$ explicitly.
These must transform according to a representation in the appropriate
conjugacy
class, and with the correct conformal dimension. In the $Spin(24)$
theory they are
formed from the $24$ $\bar{\lambda}^i$
($\bar{\Delta}=\half$; $V$ representation) and the spin fields ${\bar
S}^\alpha$ and
${\bar S}^{\dot\alpha}$ ($\bar{\Delta}=3/2$; $S$ or $C$). In the
$Spin(8)\times E_8$
theory, there are only $8$ $\bar{\lambda}^i$ and
the spin fields have dimension $\bar{\Delta}=1/2$; but then there are
also operators from the $E_8$ part, including the adjoint operator
$\bar{\cal J}^{ab}$ with $\bar{\Delta}=1$. In either theory there are
clearly numerous
ways to construct operators with appropriate representations and
dimensions.
Thus, unlike the type 0 and type II theories, here, because
the central charge of the left movers, there is a large spectrum
of physical operators obtained using the left moving oscillators.

\subsec{Compactification with $(-)^{F_L}$ twist (thermal theory)}

We next consider the twisted theory where motion around the circle is
accompanied
by action with $(-1)^{F_L}$. The spectrum at generic radius $R$
includes the currents
$\cal J$ and $\bar{\cal J}$ and the discrete states $G$ and $A^{IJ} $
of \twefourv\  as
well as
\eqn\twert{\eqalign{
  &T_V=e^{-\varphi}\bar \CO_V(\half +2nw) V_{n,2w}
  \cr
  &T_0=e^{-\varphi}\bar \CO_0(\half +(n+\half)(2w+1)) V_{n+\half,2w+1}
  \cr
  &\Psi_S = e^{-\half\varphi + i\half H }\bar \CO_S(
  \half +(2n+1)w) V_{n+\half,2w}~,\qquad ~~p_R \geq 0  \cr
  &\Psi_C = e^{-\half\varphi + i\half H }\bar \CO_C(
  \half +n(2w+1)) V_{n,2w+1}~,\qquad ~~p_R \geq 0  \cr
  &\tilde \Psi_S = e^{-\half\varphi - i\half H }\bar \CO_S(
  \half +n(2w+1)) V_{n,2w+1}~, \qquad ~~p_R \leq 0  \cr
  &\tilde \Psi_C = e^{-\half\varphi - i\half H }
  \bar \CO_C(\half +(2n+1)w) V_{n+\half,2w}~, \qquad ~p_R \leq 0 \cr
  }}
where again $n$ and $w$ are integers. The untwisted sector consists of
all
states with even winding; the spacetime bosons (fermions)
have integer (half-integer) momentum, to compensate for the action with
$(-1)^{F_L}$.
The twisted sector (odd winding) has the opposite correlation. The
spectrum \twert\ is
modular invariant
(some details of this are discussed in section 4.2).

The transformation $R\to 1/R$ leaves the set of operators \twert\
invariant,
except for the trivial interchange of the $S$ and $C$ conjugacy classes.
Hence the theory is self-dual.

Let us be explicit about the decompactification limits: as $R\to\infty$
only operators
with no winding remain, such as $\Psi_S$ and $\tilde{\Psi}_C$ . As
$R\to 0$ it is the
operators with vanishing momentum that remain, including $\Psi_C$ and
$\tilde{\Psi}_S$. The spectra in the two limits are thus related by
charge conjugation
${\cal C}$ \Cdef\ which, as discussed in section 2.3, amounts to a
change of
convention with no physical significance.

At the self-dual point $R=1$ there are additional discrete states
\eqn\enhsymm{A_{\pm}^I  =  {\cal J}{\bar\lambda^I}e^{\pm i {\bar x}} }
where $I=1,\cdots,N$ ($N=24$ for the $Spin(24)$ theory and $N=8$ for
$Spin(8)\times E_8$).
Taken together with the operators $A^{IJ}$ and $G$
from \twefourv\ these form an $Spin(N+2)$ current algebra at level $1$.
This means the left moving symmetry is enhanced from $Spin(N)\times
U(1)$
to $Spin(N+2)$ at the self-dual point.

The list \twert\ includes two states that become massless at $R=1$
\eqn\tachadd{
T_0^\pm = e^{-\varphi}V_{\pm\half,\mp 1} = e^{-\varphi} e^{\pm i{\bar
x}} e^{\phi + {\bar\phi}}
}
so that, in total, there are $N+2$ tachyons at $R=1$, transforming in
the vector of
$Spin(N+2)$. These modes are massless in the sense of one-dimensional
Liouville theory \SeibergEB, that is, they have Liouville dressing
$e^{\phi + \bar{\phi}}$. In section 4.2 we will show
that they signal a phase transition at $R=1$.

At $R=1$ the $\Psi_S$ and $\Psi_C$ combine into $Spin(N+2)$ spinors, as
do $\tilde\Psi_S$ and $\tilde\Psi_C$.

\subsec{Compactification with $(-)^{f_L}$ twist}
As the final compactification we consider the twisted theory where
motion around
the circle is accompanied by $(-)^{f_L}$. The propagating modes are
\eqn\newtwist{\eqalign{
&T_V = e^{-\varphi} \bar{\cal O}_V(\half+(2n+1)w) V_{n+{1\over 2},2w}
\cr
& T_C = e^{-\varphi} \bar{\cal O}_C (\half + n(2w+1)) V_{n,2w+1} \cr
&\Psi_S = e^{-\varphi/2 + iH/2} \bar{\cal O}_S (\half + 2nw) V_{n,2w}~,
\quad\quad\quad\quad\quad\quad~~~~~~ p_R\geq 0~~\cr
&\tilde\Psi_C =e^{-\varphi/2 - iH/2} \bar{\cal O}_C (\half + (2n+1)w)
V_{n+{1\over 2},2w}~,\quad\quad\quad
~~~ p_R\leq 0\cr
&\Psi_0 =e^{-\varphi/2 + iH/2} \bar{\cal O}_0 (\half + (n+\half)(2w+1))
V_{n+{1\over 2},2w+1}~,~~p_R\geq 0~~\cr
&\tilde\Psi_V =e^{-\varphi/2 - iH/2} \bar{\cal O}_V (\half +
n(2w+1))V_{n,2w+1}~,\quad\quad\quad
~~~p_R\leq 0
}}
The untwisted sector (even winding) has momentum shifted by half for
odd world-sheet
fermion number ($V$ and $C$ conjugacy classes). The twisted sectors
have the
opposite correlation. The theory is modular invariant (shown in detail
in section 4.3).

The list \twert\ includes two fermionic states that become massless (in
the one-dimensional
Liouville sense described below \tachadd) at $R=1$
\eqn\addpsi{
\Psi_0^{\pm} = e^{-\varphi /2 + i H /2} e^{\pm i \bar{x} } e^{\phi
+\bar{\phi}}
}
These will be responsible for a phase transition at $R=1$.

At $R=1$ the operator $\Psi_0$ with $n=w=0$ is the gravitino
$S^+{\bar{\cal J}}$
where
\eqn\susy{
{\cal S}^+ = e^{-\varphi/2 + iH/2} V_{1/2,1} = e^{-\varphi/2 + iH/2+ix}
}
is a $(1,0)$ ``supersymmetry'' current. It is a special case of the
construction of
\KutasovUA. It exists only at precisely $R=1$
because only then does $V_{1/2,1}$ have conformal dimension
$\Delta=1/2$.

In the $Spin(8) \times E_8$ (but {\it not} the $Spin(24)$ theory) there
are
also additional discrete states at $R=1$:
\eqn\twistenh{A_{\pm}^\alpha = {\cal J} \bar{S}^\alpha e^{\pm i \bar{x}}
}
These combine with $G$ and $A^{IJ}$ to extend the gauge symmetry to
$Spin(10)\times E_8$. The $\Psi_S$ and $\Psi_0$ combine into
representations
in the spinor conjugacy class of $Spin(10)$. Similarly, $T_V$ and $T_C$
combine
into $Spin(10)$ spinors, as do $\tilde\Psi_V$ and $\tilde\Psi_C$.

Next we consider duality of the theory. Transforming $R\to 1/R$ on the
operators
\newtwist\ we find that the spectrum returns to its original form {\it
except}
that the modings of the operators in $V$ and $C$ conjugacy classes have
been
interchanged. In the $Spin(8)\times E_8$ theory this is just triality,
which just amounts to a change of conventions; so this
theory is self-dual. Again the duality is an element of the enhanced
gauge symmetry
at $R=1$.

The $Spin(24)$ theory is more confusing. Formally, the $R\to 1/R$ again
interchanges the $V$ and $C$ conjugacy classes. However, for $Spin(24)$
these representations are not equivalent, nor is there enhanced gauge
symmetry at $R=1$. We therefore find that there is a whole line of
inequivalent
theories. As $R\to 0$ the spectrum degenerates to the $(-)^{f_L}$
orbifold theory
which, as discussed in the end of section 2.4, appears inconsistent.

As summary of this section, we tabulate for each line of theories the
duality
symmetry and the enhanced gauge symmetry at the self-dual point.
   \bigskip
\vbox{
$$\vbox{\offinterlineskip
\hrule height 1.1pt
\halign{&\vrule width 1.1pt#
&\strut\quad#\hfil\quad&
\vrule width 1.1pt#
&\strut\quad#\hfil\quad&
\vrule width 1.1pt#
&\strut\quad#\hfil\quad&
\vrule width 1.1pt#\cr
height3pt
&\omit&
&\omit&
&\omit&
\cr
&\hfil  &
&\hfil $Spin(24)$ &
&\hfil $Spin(8) \times E_8$ &
\cr
height3pt
&\omit&
&\omit&
&\omit&
\cr
\noalign{\hrule height 1.1pt}
height3pt
&\omit&
&\omit&
&\omit&
\cr
&\hfil $S_1$ &
&\hfil $R \to {2 \over R}$; $Spin(24)\times SU(2)$ &
&\hfil $R \to {2 \over R}$; $Spin(8) \times SU(2)\times E_8 $ &
\cr
height3pt
&\omit&
&\omit&
&\omit&
\cr
\noalign{\hrule}
height3pt
&\omit&
&\omit&
&\omit&
\cr
&\hfil $S_1 / (-1)^{F_L}$&
&\hfil $R \to {1 \over R}$; $Spin(26)$ &
&\hfil $R \to {1 \over R}$; $Spin(10) \times E_8$ &
\cr
height3pt
&\omit&
&\omit&
&\omit&
\cr
\noalign{\hrule}
height3pt
&\omit&
&\omit&
&\omit&
\cr
&\hfil $S_1 / (-1)^{f_L}$&
&\hfil No duality/enhancement&
&\hfil $R \to {1 \over R}$; $Spin(10) \times E_8$ &
\cr
height3pt
&\omit&
&\omit&
&\omit&
\cr
}\hrule height 1.1pt
}
$$
}
\centerline{\sl Table 1: Summary of T-duality symmetry and enhanced
gauge symmetry at the self-dual point.}

\newsec{The Torus Partition Functions}
In this section we analyze the torus partition function of the
compactified $Spin(24)$ and $Spin(8)\times E_8$ theories. We
consider in turn the three theories discussed above: untwisted,
thermal twist, and twist by world-sheet fermion number
$(-)^{f_L}$. The result in each case takes the form \eqn\genpart{
Z = aR +{b\over R} } for some constants $a$ and $b$. In the
twisted theories there is a phase transition at $R=1$; and so the
constants $a,b$ are different for $R>1$ and $R<1$. In each case we
compute $a,b$ and perform non-trivial checks on our results:
\item{(i)} We rewrite the string theory partition function in a
form that isolates the field theory result (proportional to $1/R$)
and the cosmological constant (proportional to $R$). The
coefficients $a,b$ are computed unambiguously this way.
\item{(ii)} The procedure in {(i)} uncovers a non-analytic
contribution to the torus partition function of the twisted
theories. This signals phase transitions at $R=1$ for all the
twisted theories. We trace the non-analytic term to modes that
become massless at $R=1$ and show how it arises in conventional
field theory. \item{(iii)} We compute the coefficient $b$
independently in field theory. As explained in
\refs{\KlebanovQA,\DouglasUP,\SeibergBX} this can be implemented
efficiently by summing over momentum modes using $\zeta$-function
regularization $\sum_{n=1}^\infty n \to -{1\over 12}$ and
$\sum_{n=0}^\infty (n+{1\over 2}) \to {1\over 24}$ for bosons
(opposite sign for fermions). In the twisted theories these
results are reliable for $R>1$ only. \item{(iv)} The coefficient
$a$ (the cosmological constant) is independent of the boundary
condition ({\it i.e.}\ insensitive to the twists). This is a
non-trivial check on the computations. Additionally, the
coefficient $a$ can be computed as in (iii), but now summing over
winding modes. When there is phase transition the result obtained
this way can be trusted only for $R<1$.

\noindent
Our final results are given in (4.8) and (4.9) for
the untwisted theory and tables 2 and 3 in the twisted cases. The
interpretation of the
results is more tentative; it is discussed in section 4.4.

A final point to make before we move on concerns the calculation of the
odd spin
structure of the right-moving CFT. In higher dimensions the odd spin
structure
vanishes trivially, due to the presence of fermion zero modes; but, in
two
dimensions, these can be cancelled and a non-zero contribution arises in
some cases \DouglasUP. For heterotic strings there are additional
fermionic
zero-modes due to the left moving fermions so, for diagonal GSO
projection,
the odd spin structure is again irrelevant. The only remaining question
is for the
compactifications with twist where, in general, the odd spin structure
multiplies
nonvanishing left movers. We now argue, following Appendix A.1 of
\DouglasUP,
that the odd spin structure in fact vanishes quite generally for
heterotic strings.

The effect of the zero mode of the superconformal ghost, $\gamma$, is
to cancel
the zero mode of the fermionic partner of the Liouville field,
$\psi_\phi$,
for both are associated
with conformal Killing spinors on the torus. The zero mode of the
supermodulus,
$\beta$, leads to an insertion into the path integral on the torus of
the supercharge
\eqn\supercharge{
G(z) = \psi_x \partial x + \psi_\phi \partial \phi - 2  \partial
\psi_\phi
}
This insertion absorbs the $\psi_x$ zero mode leaving only $\langle
\partial x(z)
\rangle$ to be calculated, where only $x$ is to be integrated
over. This vanishes due to the $x \to -x$ symmetry of the worldsheet
theory and so the odd spin structure also vanishes. In a theory with
both left- and right-moving supercharges, the final result would be
proportional to $\langle \partial x \bar{\partial} x \rangle$ and thus
generically non-zero. So we see that it is a feature of the heterotic
theories that this spin-structure vanishes even in $D=2$.

\subsec{Torus partition function: untwisted theory}
First, consider the compactified theory with no twists. The partition
function of the matter field
alone is
\eqn\zxone{\eqalign{Z_x(\tau) &=  {1\over |\eta(\tau)|^2}
\sum_{n,w\in\ZZ} q^{ {1\over 2}p^2_R}
{\bar q}^{ {1\over 2}p^2_L} \cr
&= 2\pi R \cdot {1\over\sqrt{8\pi^2 \tau_2}}~{1\over |\eta(\tau)|^2}
\sum_{m,w\in\ZZ}
\exp\left( -{\pi R^2 |m-w\tau|^2 \over 2\tau_2}\right)
}}
In the first line the lattice sum is over $p_{R,L} = {n\over R}\pm
{wR\over 2}$. The symmetry
under $R\to {2\over R}$ is manifest in this form. The second line
(obtained by Poisson
resummation) is the instanton sum which is more convenient here. The
corresponding
partition function for the Liouville field is regulated by a volume
$V_L$ and there is no sum over instantons.
The contribution from bosonic ghosts is $|\eta(\tau)|^4$. The complete
torus partition function
then takes the form
\eqn\torus{
Z_{\rm circle} = 2\pi R\cdot V_L ~\int_{\cal F} {d\tau d{\bar\tau}\over
4\tau_2}~{1\over 8\pi^2 \tau_2}
~ Z_F(\bar{\tau})~
\sum_{m,w\in\ZZ} \exp\left(-{\pi R^2 |m - w\tau|^2\over 2\tau_2}\right)
}
where, in the present section, we write
\eqn\ncpart{
Z_F  (\bar{\tau}) = \half \left( \chi_0^0 (\bar{\tau})  - \chi_1^0
(\bar{\tau}) - \chi_0^1  (\bar{\tau}) \right)
}
as a convenient expression that combines \fewterms\ and \jacid\ using
the notation
\eqn\chidef{
\chi_i^j(\bar{\tau})= \cases{
Z_i^j (\bar{\tau})^{12} & $Spin(24)$ \cr
Z_i^j (\bar{\tau})^{12}\left( Z_0^0 (\bar{\tau})^8 + Z_1^0
(\bar{\tau})^8 + Z_0^1 (\bar{\tau})^8 \right)  & $Spin(8)$}
}
The right-moving fermions and the super-conformal ghosts are included
in $Z_F$; they cancel
except for the relative signs of the various terms.

The modular integral can be carried out explicitly as follows
\refs{\PolchinskiZF\AlvarezUK\BrienPN-\GrossUB}.
First, divide the instanton sum over $(m,w)$ into the zero-mode $m=w=0$
and the
non-zero-modes. Next, rewrite the non-zero-modes
$m=kp$ and $w=kq$ with $k={\rm gcd}(m,w)$; then $p,q$ are mutually
prime.
For each mutually prime pair $p,q$ there is a
unique modular tranformation $(p,q)\to (1,0)$ that maps the fundamental
region ${\cal F}$ ($|\tau|>1$, $|\tau_1|< {1\over 2}$, $\tau_2>0$) to a
domain
$E_{p,q}\subset E$, where $E$ is the half-strip
($-{1\over 2}<\tau_1< {1\over 2}$, $\tau_2>0$). The
union $E=\cup_{p,q}E_{p,q}$ makes up the entire half-strip so
the net result
is to trade the sum over all non-zero-modes for a sum over only $(k,0)$,
while simultaneously extending the integration region from the
fundamental region ${\cal F}$ to the entire half-strip $E$. In the
present context this
procedure gives
\eqn\eval{
Z_{\rm circle} ={RV_L\over 16\pi}  ~
\left[  \int_{\cal F}{d\tau d{\bar{\tau}}\over \tau_2^2}
Z_F(\bar{\tau})+ 2 \sum^\infty_{k=1}
\int_E {d\tau d{\bar{\tau}}\over \tau_2^2} Z_F(\bar{\tau})
e^{-{\pi R^2 k^2 \over  2\tau_2}} \right]
}
The first term (integration over ${\cal F}$) gives the cosmological
constant, and the
second term (integration over $E$) gives the standard quantum field
theory result.
In the field theory term, the integral over $\tau_1$ simply implements
level matching.

In the $Spin(24)$ theory, carrying out the integral (note $d\tau d{\bar\tau}=2d\tau_1 d\tau_2$)
gives
\eqn\finpart{
Z_{{\rm circle}, Spin(24)}  = {RV_L\over 16\pi}  ~24~
  \left( {2\pi\over 3} + 4\cdot {2\over \pi R^2} \cdot {\pi^2\over 6}
\right)
= V_L\left(
R + {2\over R} \right)
  }
where we have used $Z_F = 24$. As a check, note that \finpart\ is
consistent with the self-duality $R\to 2/R$. In the $Spin(8)\times
E_8$ theory the fermionic partition function \jacid\ vanishes and
so \eqn\finpar{ Z_{{\rm circle}, Spin(8)} =0 }

We can understand the results \finpart\ and \finpar\ independently from
field theory: for $Spin(24)$ we
sum over momenta $\sum_n {n\over R}\to-{1\over 12}{1\over R}$ for each
of the $24$
spacetime bosons. Multiplication by $(-V_L)$ then gives the field
theory term in \finpart.
Similarly, summing over the winding $\sum_w {wR\over 2} \to -{1\over
24}R$ for each
of the $24$ bosons, we recover the term proportional to $R$. In the
$Spin(8)\times E_8$
theory the $8$ bosons and $8$ fermions cancel in each case; so
the vanishing partition function \finpar\ follows from field theory as
well.

\subsec{Torus partition function: thermal theory}
We next consider the theory with a $(-)^{F_L}$ twist.
The twisting correlates the left-moving fermions and the lattice vectors
non-trivially, as indicated in \twert. The torus partition function for
the thermal theory
is
\eqn\toru{\eqalign{
Z_{\rm thermal} &=  \int_{\cal F} {d\tau d{\bar\tau}\over
4\tau_2}~{V_L\over \sqrt{8\pi^2 \tau_2}}\cdot
{1\over 2} \sum_{n,w\in\ZZ} \left\{  \left[\chi^0_0(\bar{\tau}) -
\chi^0_1(\bar{\tau})\right]
q^{{1\over 2}({n\over R}+{2wR\over 2})^2}{\bar q}^{{1\over 2}(
{n\over R}-{2wR\over 2})^2}
\right. \cr
  +& \left[\chi^0_0(\bar{\tau}) + \chi^0_1(\bar{\tau})\right]
q^{ {1\over 2}({n+{1\over 2}\over R}+
{(2w+1)R\over 2})^2}{\bar q}^{{1\over 2}(
{n+{1\over 2}\over R}-{(2w+1)R\over 2})^2} \cr -& \left.
\chi^1_0(\bar{\tau})\left[q^{{1\over 2}({n+{1\over 2}\over R}+
{2wR\over 2})^2}{\bar q}^{{1\over 2}(
{n+{1\over 2}\over R}-{2wR\over 2})^2} +
q^{{1\over 2}({n\over R}+
{(2w+1)R\over 2})^2}{\bar q}^{{1\over 2}(
{n\over R}-{(2w+1)R\over 2})^2} \right]
   \right\}
}}
This expression is absolutely convergent for all $R$ but, due to the
additional tachyons
$T_0^\pm$ \tachadd\ at $R=1$, it is {\it not} analytic in $R$.  To see
this explicitly, focus on the
contribution from these states:
\eqn\zsing{
Z_{\rm thermal} = \int {d\tau d{\bar\tau}\over 4\tau_2} {V_L\over
\sqrt{8\pi^2\tau_2}}~
{1\over 2} \left[ \chi_0^0(\bar{\tau}) +
\chi_1^0(\bar{\tau})\right]\cdot
2{\bar q}^{{1\over 8}(R+{1\over R})^2} q^{{1\over 8}(R-{1\over R})^2} +
{\rm regular}
}
Since ${1\over 2} ( \chi_0^0(\bar{\tau}) + \chi_1^0(\bar{\tau}))\sim
{\bar q}^{-1/2}$ for large $\tau_2$
the exponential damping disappears for $R=1$; the integral remains
convergent at $R=1$
only because of the powers of $\tau_2$. However, the second derivative
$\partial^2_RZ_{\rm th}$ diverges
at $R=1$. This establishes a first order phase transition at $R=1$.

The expression \toru\ is symmetric under the duality $R\to 1/R$.
However, because
of the phase transition, we focus for now on $R>1$.
Poisson resummation on the momenta gives
  \eqn\torus{\eqalign{
  Z_{\rm thermal} &=  \int_{\cal F} {d\tau d{\bar\tau}\over 4\tau_2}
  ~{2\pi R\cdot V_L\over 8\pi^2 \tau_2}\cdot
  {1\over 2} \sum_{m,w\in\ZZ}
  \left\{ \left[\chi^0_0(\bar{\tau}) - \chi^0_1(\bar{\tau})
  -(-)^m \chi^1_0(\bar{\tau}) \right]e^{-S(m,2w)} \right. \cr
  +&  \left.  \left[(\chi^0_0(\bar{\tau}) + \chi^0_1(\bar{\tau}))
  (-)^m- \chi^1_0(\bar{\tau}) \right] e^{-S(m,2w+1)}\right\}
  }}
where $S(m,w) = {\pi R^2\over 2\tau_2} |m - w\tau|^2$. For
sufficiently small $R$, the integral over individual terms in
\torus\ diverges. The finite answer depends on first performing
the sum over $m,w$ and then integrating over $\tau$. This lack of
absolute convergence eventually leads to our phase transition.

It is useful to write \torus\ as \eqn\zhetintro{Z_{\rm thermal}=
Z_{\rm circle} -2Z_{\rm flip} } where we assembled the ``wrong
sign" contributions into \eqn\zhetdef{\eqalign{ Z_{\rm flip} &=
{RV_L\over 16\pi} \int_{\cal F} {d\tau d{\bar\tau}\over\tau_2^2}~
{1\over 2} \sum_{m,w\in\ZZ} \left[ \chi^0_0(\bar{\tau})
e^{-S(2m+1,2w+1)}
  -\chi^0_1(\bar{\tau}) e^{-S(2m,2w+1)} \right. \cr
&\quad \left. - \chi^1_0(\bar{\tau}) e^{-S(2m+1,2w)}
  \right]
}}
This expression is modular invariant:\  modular transformations
permute the
instanton factors $S(2m+1,2w+1)$, $S(2m,2w+1)$,
and $S(2m+1,2w)$ nontrivially, but the fermion factors
$\chi^0_0$, $-\chi^0_1$, and $-\chi^1_0$ compensate for this, because
they are
permuted in exactly the same way. When mapping to the half-strip, the
three instanton
terms all map into $(2k+1)^2 S(1,0)$ (the greatest common divisor is
odd in each case)
and the fermion partition factors in each case maps to the $-\chi^1_0$.
Between the
three terms, all pairs of mutually primes are being covered; so the
union of the integration
regions after mapping is again the entire half-strip $E$. Therefore,
the integral \zhetdef\ can
be written as
\eqn\comzhet{
Z_{\rm flip} = - {RV_L\over 16\pi}
\int_E {d\tau d{\bar\tau}\over\tau_2^2}~
{1\over 2} \chi^1_0(\bar{\tau})  \sum_{k\in\ZZ} e^{-(2k+1)^2 S(1,0)}
}
Collecting terms we find
  \eqn\torusth{\eqalign{
  Z_{\rm thermal} &= {RV_L\over 16\pi} \left\{  \int_{\cal F}
  {d\tau d{\bar\tau}\over \tau^2_2}~{1\over 2}
  \left[\chi^0_0(\bar{\tau}) - \chi^0_1(\bar{\tau})
  -\chi^1_0(\bar{\tau}) \right] \right. \cr
  +&  \left. 2\sum_{k=1}^\infty \int_E {d\tau d{\bar\tau}
  \over \tau^2_2}~ {1\over 2}  \left[\chi^0_0(\bar{\tau}) -
  \chi^0_1(\bar{\tau}) -(-)^k \chi^1_0(\bar{\tau}) \right]e^{-S(k,0)}
  \right\}
  }}
The significance of $Z_{\rm flip}$ in \torus\ is, therefore, to
reverse the sign of the field theory contribution from spacetime
fermions with odd momentum, as expected of thermal twisting.

In the form \torusth\ the connection to standard thermodynamics is
clear \PolchinskiZF.  The first term is the vacuum energy and the
second term is the trace over the spacetime Hilbert space of
$e^{-\beta H}$. In the second term of \torusth\ one has to first
integrate over $\tau_1$ thus implementing the level matching
condition in the theory in noncompact space, and then integrates
over $\tau_2$. In our case this integral over $\tau_2 $ converges.
However, recall the observation after \torus\ that, for sufficiently
small $R$, the integral over $\tau$ of the individual terms in the sum
diverges, and the correct, finite answer is obtained only if one first
performs the sum and then integrate over $\tau$.  In the form \torusth\ this
translates to the statement that the result of the integral over
the half strip $E$ depends on the precise way it is performed. In
particular, the correct answer might not correspond to first
integrating over $\tau_1$ and then over $\tau_2$.  We will now see
that for small $R$ this naive result of \torusth\ is in fact
wrong.

In the $Spin(24)$ theory there are no spacetime fermions;
$Z_{\rm flip}=0$ by level matching. Accordingly, the torus
partition function  \torusth\ for the thermal theory agrees
exactly with the result \finpart\ from the circle theory
  \eqn\affz{\eqalign{
  Z_{{\rm thermal}, Spin(24)} &= V_L \left( {R} + {2\over R}
\right)\quad\quad,~R>1
  }}
This makes sense physically: the cosmological constant term
(proportional to $R$) should not
be sensitive to boundary conditions; and the field theory contribution
(proportional to $1/R$)
of the $24$ spacetime bosons is also unaffected by the twist.

The result for $R<1$ is easily determined as
\eqn\affhighT{\eqalign{
  Z_{{\rm thermal}, Spin(24)} &= V_L \left( 2 R+ {1\over R}
\right)\quad\quad,~R<1
}}
by imposing the symmetry under $R\to 1/R$ of the original expression
\toru. In more
detail, this can be obtained by Poisson resummation on $w$, rather than
$n$. This
gives alternate expressions similar to \torus, \zhetdef, and \torusth;
but now with good
convergence for small $R$.

The two expressions  \affz\ and \affhighT\ agree at $R=1$ but, as
expected from \zsing,
the derivative is discontinuous there. This result can be understood
from effective
field theory, as explained in the introduction. Indeed,
combining the string theory results \affz\ and \affhighT\ into
\eqn\zthallT{
Z_{{\rm thermal}, Spin(24)} = {3\over 2} V_L\left( R + {1\over
R}\right) -
{1\over 2}V_L\left| R - {1\over R}\right| \quad\quad,~\forall R
}
we see that the field theory result \lgonel\ for a complex boson with
$m(R)=\half \left|R-{1 \over R}\right|$ accounts precisely for the
non-analyticity.

For the $Spin(8)\times E_8$ theory the integral \torusth\ gives
\eqn\affztwo{
  Z_{{\rm thermal}, Spin(8)} ={V_L \over R }\quad\quad,~R>1
  }
The cosmological constant (proportional to $R$) vanishes as it did for
the circle theory;
so it is independent of the twist as it should be. The  field theory
contribution is
nonvanishing because the spacetime fermions are sensitive to the
$(-)^k$ weight of the
instanton sectors in \torusth; they no longer cancel the bosons.
The result \affztwo\ also follows from field theory, by summing up the
momenta $\sum_n n/R$ of $8$ bosons and the shifted momenta
$\sum_n (n+{1\over 2})/R$ of $8$ fermions.

The partition function in the small $R$ phase follows from duality:
  \eqn\affzto{
  Z_{{\rm thermal}, Spin(8)} = V_L {R}\quad\quad,~R<1
  }
Combining the results \affztwo\ and \affzto\ into
\eqn\zthallT{
Z_{{\rm thermal}, Spin(8)} = {1\over 2} V_L\left( R + {1\over R}\right)
-
{1\over 2}V_L\left| R - {1\over R}\right| \quad\quad,~\forall R
}
we see that the non-analyticity in the $Spin(8)\times E_8$ theory is
identical to that found
in the $Spin(24)$ theory; it is again accounted for by the complex
boson with
$m(R)=\half \left|R-{1 \over R}\right|$. For easy reference we
summarize the results for the thermal
theory as follows:
\bigskip
\vbox{
$$\vbox{\offinterlineskip
\hrule height 1.1pt
\halign{&\vrule width 1.1pt#
&\strut\quad#\hfil\quad&
\vrule width 1.1pt#
&\strut\quad#\hfil\quad&
\vrule width 1.1pt#
&\strut\quad#\hfil\quad&
\vrule width 1.1pt#\cr
height3pt
&\omit&
&\omit&
&\omit&
\cr
&\hfil &
&\hfil $R>1$&
&\hfil $R<1$  &
\cr
height3pt
&\omit&
&\omit&
&\omit&
\cr
\noalign{\hrule height 1.1pt}
height3pt
&\omit&
&\omit&
&\omit&
\cr
&\hfil $Z_{thermal,Spin(24)}$&
&\hfil $V_L \left(R + {2\over R} \right)$ &
&\hfil $V_L\left(2R + {1\over R} \right)$ &
\cr
height3pt
&\omit&
&\omit&
&\omit&
\cr
\noalign{\hrule}
height3pt
&\omit&
&\omit&
&\omit&
\cr
&\hfil $Z_{thermal,Spin(8)}$&
&\hfil $V_L  {1\over R} $ &
&\hfil  $V_L R$ &
\cr
height3pt
&\omit&
&\omit&
&\omit&
\cr
}\hrule height 1.1pt
}
$$
}
\centerline{\sl Table 2: Torus partition functions after the thermal
twist $(-)^{F_L}$. }

\subsec{Torus partition function: theory with $(-)^{f_L}$ twist}
The theory with a $(-)^{f_L}$ twist has the partition function
\eqn\wsd{
\eqalign{
Z_{\rm twist} & = \int_{\cal F} {d\tau d{\bar\tau}\over
4\tau_2}~{V_L\over \sqrt{8\pi^2 \tau_2}}
\sum_{n,w\in\ZZ} \left\{{1\over 2} ( \chi_0^0(\bar{\tau}) -
\chi_1^0(\bar{\tau}))
q^{{1\over 2}({n+{1\over 2}\over R}+ wR)^2}
{\bar q}^{{1\over 2}({n+{1\over 2}\over R}- wR)^2} \right. \cr
&-{1\over 4} \chi_0^1(\bar{\tau}) (q^{{1\over 2}({n\over R}+
wR)^2}{\bar q}^{{1\over 2}({n\over R}- wR)^2}
+q^{{1\over 2}({n+{1\over 2}\over R}+ wR)^2}
{\bar q}^{{1\over 2}({n+{1\over 2}\over R}- wR)^2})\cr
&+{1\over 2} \chi_0^1(\bar{\tau}) q^{{1\over 2}({n\over R}+ {2w+1\over
2}R)^2}
{\bar q}^{{1\over 2}({n\over R}-{2w+1\over 2}R)^2}\cr
& -{1\over 4} ( \chi_0^0(\bar{\tau}) + \chi_1^0(\bar{\tau}))q^{{1\over
2}({n+{1\over 2}\over R}+
{2w+1\over 2}R)^2}
{\bar q}^{{1\over 2}({n+{1\over 2}\over R}-{2w+1\over 2}R)^2}
\cr
& \left.  - {1\over 4} ( \chi_0^0(\bar{\tau}) -
\chi_1^0(\bar{\tau}))q^{{1\over 2}({n\over R}+ {2w+1\over 2}R)^2}
{\bar q}^{{1\over 2}({n\over R}-{2w+1\over 2}R)^2} \right\}
}}
As in the thermal theory, there is a non-analytic feature at $R=1$,
interpreted as a phase
transition. The origin of the phase transition is the two fermions
\addpsi\ that become
massless at $R=1$. Their contribution to \wsd\ is
\eqn\divterm{
Z_{\rm twist} = \int_{\cal F} {d\tau d{\bar\tau}\over
4\tau_2}~{V_L\over \sqrt{8\pi^2 \tau_2}}
(-{1\over 4})( \chi_0^0(\bar{\tau}) +
\chi_1^0(\bar{\tau}))~2~q^{{1\over 8}({1\over R}+
R)^2}
{\bar q}^{{1\over 8}({1\over R}-R)^2} + {\rm regular}
}
which clearly has a divergent second derivative.

After Poisson resummation, the partition function can be written as
\eqn\ws{
\eqalign{
Z_{\rm twist} & = {R V_L \over 16 \pi} \int_{\cal F} { d\tau
d{\bar{\tau}} \over \tau_2^2} \sum_{m,w\in\ZZ} \left\{ {1\over 2}
\chi_0^0(\bar{\tau}) (e^{-S(2m,2w)} -
e^{-S(2m+1,2w)}- e^{-S(2m,2w+1)}) \right. \cr
&- {1\over 2} \chi_1^0(\bar{\tau}) (e^{-S(2m,2w)}- e^{-S(2m+1,2w+1)} -
e^{-S(2m+1,2w)}) \cr
& \left.  -{1\over 2} \chi_0^1(\bar{\tau}) (e^{-S(2m,2w)}-
e^{-S(2m,2w+1)} - e^{-S(2m+1,2w+1)}) \right\}
}}
This expression is manifestly modular invariant. Comparing with
quantities defined in
the previous sections, we find
\eqn\relpart{
Z_{\rm twist}(R) = Z_{\rm circle}(2R) - {1\over 2} Z_{\rm circle}(R)
-{1\over 2} Z_{\rm thermal}(R)
}
for all $R$. This gives
\bigskip
\vbox{
$$\vbox{\offinterlineskip
\hrule height 1.1pt
\halign{&\vrule width 1.1pt#
&\strut\quad#\hfil\quad&
\vrule width 1.1pt#
&\strut\quad#\hfil\quad&
\vrule width 1.1pt#
&\strut\quad#\hfil\quad&
\vrule width 1.1pt#\cr
height3pt
&\omit&
&\omit&
&\omit&
\cr
&\hfil &
&\hfil $R>1$&
&\hfil $R<1$  &
\cr
height3pt
&\omit&
&\omit&
&\omit&
\cr
\noalign{\hrule height 1.1pt}
height3pt
&\omit&
&\omit&
&\omit&
\cr
&\hfil $Z_{twist,Spin(24)}$&
&\hfil $V_L \left(R - {1\over R} \right)$ &
&\hfil ${V_L\over 2} \left(R - {1\over R} \right)$ &
\cr
height3pt
&\omit&
&\omit&
&\omit&
\cr
\noalign{\hrule}
height3pt
&\omit&
&\omit&
&\omit&
\cr
&\hfil $Z_{twist,Spin(8)}$&
&\hfil -$V_L  {1\over 2R} $ &
&\hfil  -$V_L {R\over 2}$ &
\cr
height3pt
&\omit&
&\omit&
&\omit&
\cr
}\hrule height 1.1pt
}
$$
}
\centerline{\sl Table 3: Torus partition functions after twist by
$(-)^{f_L}$. }

For $R>1$ the entries can be understood from general principles: the
cosmological
constant (proportional to $R$) is the same as in the untwisted theories
\finpart,\finpar.
For $Spin(24)$
the field theory term (proportional to $1/R$) is due to $24$ bosons
with momenta
$(n+{1\over 2})/R$; and, for $Spin(8)\times E_8$ it is due
to $8$ bosons with momenta $(n+{1\over 2})/R$, $8$ real fermions with
momenta
$n/R$, and $8$ real fermions with momenta $(n+{1\over 2})/R$.

The results in the table can be combined as
\eqn\zwsall{\eqalign{
Z_{{\rm twist}, Spin(24)} &= {3\over 4} V_L\left( R - {1\over R}\right)
+
{1\over 4}V_L\left| R - {1\over R}\right| \quad\quad,~\forall R\cr
Z_{{\rm twist}, Spin(8)} &= - {1\over 4} V_L\left( R + {1\over
R}\right) +
{1\over 4}V_L\left| R - {1\over R}\right| \quad\quad,~\forall R\cr
}}
We see that, as expected, the singularity in each theory takes the form
predicted by
field theory \lgonel\ for one complex fermion with mass
$m(R)={1\over 2}\left| R - {1\over R}\right|$.
It is also manifest in \zwsall\ that only the $Spin(8)\times E_8$
theory satisfies the duality
symmetry $R\to 1/R$.

\subsec{Discussion} Up to this point we have discussed the torus
partition function as a rather abstract object. Our determination
of this object is unambiguous, but the interpretation is not
necessarily straightforward. Here we discuss these issues.
\bigskip

\centerline{\it Thermodynamic interpretation of theories twisted
by $(-)^{F_L}$.} In the thermal theory it is natural to try to
identify $2\pi R=\beta=T^{-1}$. Recalling that the total partition
function, including disconnected amplitudes, is related to the
torus amplitude through $Z_{\rm tot} = e^{Z_{\rm sphere} + Z_{\rm
torus}+...}$, we then find the energy density $\epsilon = -{1\over
V_L}{\partial\ln Z_{\rm tot}\over\partial\beta}=-{1\over V_L}
{\partial Z_{\rm torus} \over\partial\beta}+...$, and the free energy
density $f = \epsilon-Ts= -{1\over V_L\beta}\ln Z_{\rm tot}=
-{1\over V_L\beta}Z_{\rm torus}+...$. This procedure gives
  \bigskip
\vbox{
$$\vbox{\offinterlineskip
\hrule height 1.1pt
\halign{&\vrule width 1.1pt#
&\strut\quad#\hfil\quad&
\vrule width 1.1pt#
&\strut\quad#\hfil\quad&
\vrule width 1.1pt#
&\strut\quad#\hfil\quad&
\vrule width 1.1pt#\cr
height3pt
&\omit&
&\omit&
&\omit&
\cr
&\hfil &
&\hfil $T<{1 \over 2\pi}$&
&\hfil $T>{1 \over 2\pi}$ &
\cr
height3pt
&\omit&
&\omit&
&\omit&
\cr
\noalign{\hrule height 1.1pt}
height3pt
&\omit&
&\omit&
&\omit&
\cr
&\hfil $f_{24}$&
&\hfil $-4 \pi T^2  -{1 \over 2\pi}$ &
&\hfil $- 2 \pi  T^2-{1 \over \pi} $ &
\cr
height3pt
&\omit&
&\omit&
&\omit&
\cr
\noalign{\hrule}
height3pt
&\omit&
&\omit&
&\omit&
\cr
&\hfil $\epsilon_{24}$&
&\hfil $4 \pi T^2 -{ 1 \over 2 \pi}$&
&\hfil $2 \pi T^2 -{1 \over \pi}$ &
\cr
height3pt
&\omit&
&\omit&
&\omit&
\cr
\noalign{\hrule}
height3pt
&\omit&
&\omit&
&\omit&
\cr
&\hfil $s_{24}$&
&\hfil $8 \pi T$&
&\hfil $4 \pi T$&
\cr
\noalign{\hrule height 1.1pt}
height3pt
&\omit&
&\omit&
&\omit&
\cr
&\hfil $f_8$&
&\hfil $- 2 \pi T^2$ &
&\hfil $-{1 \over 2\pi} $ &
\cr
height3pt
&\omit&
&\omit&
&\omit&
\cr
\noalign{\hrule}
height3pt
&\omit&
&\omit&
&\omit&
\cr
&\hfil $\epsilon_8$&
&\hfil $ 2\pi T^2$&
&\hfil $-{1 \over 2\pi}$ &
\cr
height3pt
&\omit&
&\omit&
&\omit&
\cr
\noalign{\hrule}
height3pt
&\omit&
&\omit&
&\omit&
\cr
&\hfil $s_8$&
&\hfil $4 \pi T$&
&\hfil $0$&
\cr
height3pt
&\omit&
&\omit&
&\omit&
\cr
}\hrule height 1.1pt
}
$$
}
\centerline{\sl Table 4: Thermodynamics of the theories on a thermal
circle.}

In either theory the transition is characterized by negative latent
heat per volume
$\ell = T\Delta s=  -{1\over \pi}$. This means the high temperature
phase
is {\it more} ordered than the low temperature phase. This situation is
usually considered unacceptable in thermodynamics. Let us make some
comments on the possible interpretation of the result.

A clear benchmark for the interpretation is the evaluation of  the
trace ${\rm Tr}~e^{-\beta H}$
over the spacetime Hilbert space.
This is represented by the half-strip in our computations and gives the
result
indicated in the table for $T<{1\over 2\pi}$, but now at all $T$. The
reason this
cannot be the correct answer at high $T$ is that it is inconsistent
with the duality $R\to 1/R$;
and also it does not take into account the presence of additional light
modes
at $T={1\over 2\pi}$. However, it would be surprising if this result
was invalid for $T<{1\over 2\pi}$.

The table above does not reflect standard thermodynamics. Usually,
when considering a first order phase transition, one would reason
that near the transition there are two candidate phases which have
free energies taking the forms given in the table, with the range
for {\it each} phase analytically continued to other values of
$T$. Then one determines the stable phase as the one with the
lowest free energy at each $T$. We were guided instead by the
field theory interpretation at low temperature, and then applied
duality. Our procedure apparently amounts to taking the {\it
highest} free energy in each phase. If taken at face value this
means that both phases are unstable. However, as explained above
such an instability would be very surprising, at least in the low
temperature phase.

We are thus lead to a picture where our results and their thermodynamic
interpretation
can be trusted at $T<{1\over 2\pi}$. For $R<1$ the theories exist and
we can compute
reliably; but it seems that the proper interpretation of the string
calculation in this regime
cannot be standard thermodynamics. One indication of this is that the
torus partition
function has no obvious field theory interpretation for $R<1$.

Finally, we should clarify that, because of T-duality, the theory
with $R<1$ can, of course, be interpreted in terms of a system
with ${1\over R}>1$ and, in these variables, there exist standard
thermodynamics with temperature ${R \over 2\pi} <{1\over 2\pi}$.
The question discussed here is whether, in addition, this regime
permits an interpretation as a genuinely new phase with $T>{1\over
2\pi}$.

\bigskip

\centerline{\it Interpretation of theories twisted by $(-)^{f_L}$.}

We next discuss the interpretation of the phase transition in the
theories twisted by
$(-)^{f_L}$. Again, the results for $R>1$ have clear field theory
interpretations.
As for the regime $R<1$, a distinction must be made between the
$Spin(24)$
theory and the $Spin(8)\times E_8$ theory. In the latter, the duality
$R\to 1/R$ is
part of the gauge symmetry at the self-dual point. This means that
the $R<1$
phase
exists in the $Spin(8)\times E_8$ theory. In contrast, in the
$Spin(24)$ theory, there is
no duality $R\to 1/R$, and also no enhanced gauge symmetry at $R=1$.
In this case it is possible that the theory simply does not exist at
$R<1$.
One appealing consequence of this possibility is that it would exclude
the apparently
inconsistent limit $R\to 0$ from moduli space.

\bigskip
\noindent {\bf Acknowledgements:} \medskip \noindent

We thank O. Aharony, N. Itzhaki, D. Kutasov, J. Maldacena, E.
Martinec, S. Minwalla, and E. Witten for discussion. The work of
JLD and FL was supported in part by the DoE. The work of N.S. was
supported in part by DOE grant \#DE-FG02-90ER40542.

\appendix{A}{Lattice Constructions}
In this appendix we first classify the possible uncompactified theories
using covariant lattices,
finding exactly the $Spin(24)$ and $Spin(8)\times E_8$ theories. Next,
we consider the
compactified theory and show that there is a single contiguous moduli
space with
$13$ dimensions.
The relation between the $Spin(24)$ and $Spin(8)\times E_8$ theories is
found explicitly.

\subsec{Classification of Uncompactified Theories}
We will use the covariant lattice approach
following \LustTJ . In this formalism the superconformal ghosts are
represented
as three bosons $\vec{x}_{\rm gh}$ with canonical normalization and
signature.
The correspondence is
\eqn\vdef{
e^{q\phi} \leftrightarrow e^{i\vec{v}\cdot \vec{x}_{\rm gh}}
}
The canonical qhost pictures map as
\eqn\canpict{
\eqalign{
R:~~~~~~~ q_0 &= - {1\over 2} \leftrightarrow  \vec{v}_0 = ({1\over 2},
{1\over 2}, -{1\over 2})
\cr
NS:~~~~~~~ q_0 &= - 1 ~\leftrightarrow  \vec{v}_0 = (0 , 0 , -1 )
}}
Under this identification the levels of simple operators map as
\eqn\level{
\Delta( e^{q\phi})  = -q - {1\over 2}q^2 = {1\over 2}\vec{v}^2 = \Delta
(e^{i\vec{v}\cdot \vec{x}_{\rm gh}})}
and locality conditions are preserved because
\eqn\loc{
- q_1 q_2 = \vec{v}_1 \cdot \vec{v}_2 ~~~{\rm mod}~ 1
}
for any pairs $q_1$ and $q_2$; and the corresponding $\vec{v}_1$ and
$\vec{v}_2$.
We can now write vertex operators of propagating states as
\eqn\vert{
V_{\vec{w}_R,\vec{w}_L,k} =
e^{i\vec{w}_R \cdot \vec{H}_R }e^{i\vec{w}_L \cdot \vec{H}_L } {\cal
O}_{R,L} V_k
}
Here the $\vec{H}_R = (H,\vec{x}_{\rm gh})$ denote the $4$ right-moving
bosons
and the $\vec{H}_L$ are the corresponding $12$ bosons, making up the
lattice on the
left side. The $V_k$ is the $({1\over 2},{1\over 2})$ operator \vldef\
associated with the bosonic
matter fields $x$ and $\phi$. Finally, the ${\cal O}_{R,L}$ are
operators constructed from
the towers of bosonic oscillators. In the present context physical
conditions will not leave
any operators of this kind before compactification. Discrete states
take a similar form, with $V_k$
replaced by the identity operator.

The consistency conditions on the string theory are satisfied in the
covariant lattice construction by demanding that $w= (\vec{w}_R,
\vec{w}_L)$
forms an {\it even, self-dual lattice of signature} $(4,12)$. This
comes as
usual from the level matching condition
\eqn\level{
{1\over 2} \vec{w}^2_R - {1\over 2} \vec{w}^2_L \in \ZZ
}
which requires the lattice to be even, and invariance under the modular
transformation $\tau\to -1/\tau$ imposes self-duality. The locality
condition
\eqn\loc{ \vec{w}_{R1}\cdot \vec{w}_{R2}  - \vec{w}_{L1}\cdot
\vec{w}_{L2} \in \ZZ}
is automatic after level matching. The only subtlety in these
statements is that,
in this formalism, the description of the superconformal ghosts has
more redundancy
than is familiar, a feature that can be factored out \LustTJ. In
summary, the
uncompactified theories are classified by the even self-dual lattices
of signature
$(4,12)$.

Such lattices are in one-to-one correspondance with even self-dual
lattices in
$16$ {\it Euclidean} dimensions. It is well-known that there are
precisely
two such lattices,
$Spin(32)/\ZZ_2$ and $E_8\times E_8$. Physical string theories then
follow from
the decomposition\foot{We do not make any distinctions
between the $Spin(2n)$ lattice and the lattice of the Lie algebra
$D_n$.}
\eqn\lattsup{\Gamma_{16} \supset Spin(8) \otimes \Gamma_{12} }
where the first factor encodes the right-moving fermions and the
super-conformal ghosts,
while the second factor is the $12$-dimensional lattice of left-moving
bosons.
For $Spin(32)/\ZZ_2$ this decomposition leaves as stabilizer the
lattice $Spin(24)$
and so the first of the 2D heterotic theories. For $E_8\times E_8$ the
embedding
goes into one $E_8$-factor as $Spin(8)\subset Spin(16)\subset E_8$. The
stabilizer of this
embedding is $Spin(8)\times E_8$, leading to the other 2D heterotic
string.

In both cases the lattice embeddings align conjugacy classes of
$Spin(2n)\times Spin(2m)\subset Spin(2n+2m)$ in the obvious diagonal
fashion\foot{The
covariant lattice $0\oplus S$ gives a more symmetric decomposition.
This corresponds
to the theory obtained from ours using the ${\cal C}$ transformation
\Cdef\ which amounts
to a different convention.}
$(0,0)\oplus (V,V)\oplus (S,C) \oplus (C,S) = 0 \oplus C$ . This is the
lattice
analogue of the diagonal GSO in the CFT language.

The covariant lattice construction similarly classifies heterotic
string theories in
$10$ dimensions. In this case the relevant lattices have signature
$(8,16)$, with
the first factor $8=5+3$ from bosonized fermions and superconformal
ghosts.
Such lattices are classified by the even self-dual lattices in $24$
Euclidean dimensions,
{\it i.e.} the Niemeyer lattices. Decomposing these lattices as
$\Gamma_{24}\supset Spin(16)\otimes \Gamma_{16}$ identifies $8$
distinct heterotic string
theories  associated with different
$\Gamma_{16}$. \foot{The covariant lattice construction misses one 10D
heterotic string
theory, the one with gauge group $E_8$. The failure in this case is that
the fermions cannot be bosonized using lattices. There seems to be
no analogous possibilities in two dimensions.}
Among these, two correspond to simple factorization
$\Gamma_{24} =  Spin(16)\otimes \Gamma_{16}$. These are the usual
supersymmetric
heterotic string theories with gauge groups $Spin(32)/\ZZ_2$ and
$E_8\times E_8$,
constructed using a chiral GSO projection. In two dimensions there are
no analogues
of these theories. The remaining heterotic string theories in ten
dimensions
involve nontrivial embeddings and correspond to non-chiral GSO
projections.
The closest analogue to the 2D heterotic string theories we study
are the non-supersymmetric $Spin(32)$ theory (based on the Niemeyer
lattice $Spin(48)/\ZZ_2$)
and the $E_8\times Spin(16)$ theory (based on the Niemeyer lattice
$E_8\times Spin(32)$)
which both correspond to simple diagonal embeddings
\refs{\DixonIZ,\SeibergBY}.
The famous tachyon-free
$O(16)\times O(16)$ string theory \refs{\DixonIZ,\AlvarezGaumeJB}
is based on the Niemeyer lattice $Spin(16)^3$ which has no analogue in
$16$ Euclidean
dimensions so, from this perspective, there can be no analogous
construction in
two dimensions.

\subsec{Compactification}
We next discuss toroidal compactifications using lattices. Thus, the
matter field $X$ is
assumed periodic with period $R$. Additionally, a general
compactification
has Wilson lines. These can can be introduced as usual through the
shifted momenta
\eqn\momshift{
\eqalign{
p_R &= \left( {n\over R} - \vec{w}_L \cdot \vec{A} +{wR\over 2}
\vec{A}^2\right)
+{wR\over 2}  \cr
p_L &= \left( {n\over R} - \vec{w}_L \cdot \vec{A} + {wR\over 2}
\vec{A}^2\right)
- {wR\over 2} \cr
\vec{k_L} &= \vec{w_L} - wR\vec{A}
  }}
Here $\vec{w}_L$ refer to the vectors of the $12$ dimensional left
moving lattice
prior to compactification. Since we are using a non-standard GSO
projection,
the bosonic lattice does not decouple completely from the right moving
fermions the
way it usually does in compactifications of 10D heterotic strings. The
reason that the
usual procedure works anyway is that
\eqn\shiftGSO{ \vec{k}_L^2 + p^2_L - p^2_R = \vec{w}_L^2 - 2nw\in 2\ZZ}
so, if the original set of lattice vectors were even, then the deformed
set
is even as well as well. Additionally, if we keep the original
conjugacy classes,
the covariant lattice remains self-dual. Thus the theory must be
consistent also after
deformation.

In the covariant lattice approach, the theory is consistent exactly
when the full lattice
vector $(\vec{k}_R,p_R; \vec{k}_L, p_L)$ belongs to an even, self-dual
lattice
of signature $(5,13)$. The sublattice obtained by restricting
$\vec{k}_R$ to the
canonical ghost pictures (the $\vec{k}_R\equiv\vec{w}_R$ is unchanged
by the Wilson lines)
has signature $(1,13)$ and its moduli space is
\eqn\modspace{{\cal H}\backslash O(1,13,\RR)/O(13,\RR)
}
This $13$ dimensional moduli space of compactifications is parametrized
locally
by the radius of compactification $R$ and the $12$ Wilson lines
$\vec{A}$. The global
identifications indicated by ${\cal H}$ would be ${\cal H}=O(1,13,\ZZ)$
in standard
heterotic theory but the situation is not clear here.

Interestingly, this discussion is independent of which theory is taken
as starting point: $Spin(24)$ or $Spin(8)\times E_8$; whether twisted
or not.
This means all these theories must belong to the same moduli space;
they must
be continuously related. In the following we verify this by explicit
comparison.

Let us first show how the theory twisted by $(-)^{F_L}$, {\it i.e.} the
thermal theory,
can be obtained from the untwisted theories by turning on a suitable
Wilson line.
The computation works the same way for the $Spin(24)$ and the
$Spin(8)\times E_8$
theories so we just consider the former.

Before twisting the covariant lattice $Spin(8,24)$
can be in $0\oplus C$ which, in canonical ghost picture,
decomposes under $Spin(8)\times Spin(24)$
as $(0,0)\oplus(V,V)\oplus (C,S)\oplus (S,C)$. Each sector allows
$n,w\in\ZZ$ along the
thermal direction. The thermal twist corresponds to the Wilson line
$R\vec{A} = (1,0^{11})\in V$ of
$Spin(24)$. The shift
$\vec{w}_L\to \vec{k}_L = \vec{w}_L - w\vec{A}R$ from \momshift\  means
we still have
the conjugacy classes $(0,0)\oplus(V,V)\oplus (C,S)\oplus (S,C)$ for
$w\in 2\ZZ$ (untwisted
sector) but now the conjugacy classes are $(0,V)\oplus(V,0)\oplus
(S,S)\oplus (C,C)$ for $w\in 2\ZZ+1$
(twisted sector). The shifts \momshift\ of $p_L,p_R$ due to the Wilson
line introduces the
shifted momentum
\eqn\nshift{
\tilde{n} = n  - \vec{w}_L \cdot\vec{A}R + {w\over 2} \vec{A}^2 R^2
= n - {w\over 2} -  \vec{k}_L \cdot\vec{A}R
}
The $\vec{k}_L \cdot\vec{A}R$ is integer (half-integer) for
$\vec{k}_L\in 0\oplus V$ ($S\oplus C$) so the $\tilde{n}$ is shifted to
half-integer values
for $\vec{k_L}\in S\oplus C$ in the untwisted sector and for
$\vec{k}_L\in 0\oplus V$ in the
twisted sector. In summary, the spectrum after twisting
correlates the sectors so that
\bigskip
\vbox{
$$\vbox{\offinterlineskip
\hrule height 1.1pt
\halign{&\vrule width 1.1pt#
&\strut\quad#\hfil\quad&
\vrule width 1.1pt#
&\strut\quad#\hfil\quad&
\vrule width 1.1pt#
&\strut\quad#\hfil\quad&
\vrule width 1.1pt#\cr
height3pt
&\omit&
&\omit&
&\omit&
\cr
&\hfil $\tilde{n}$&
&\hfil $w$&
&\hfil $Spin(8,24)$&
\cr
height3pt
&\omit&
&\omit&
&\omit&
\cr
\noalign{\hrule height 1.1pt}
height3pt
&\omit&
&\omit&
&\omit&
\cr
&\hfil $\ZZ$&
&\hfil $2\ZZ$&
&\hfil $(0,0),(V,V)$ &
\cr
height3pt
&\omit&
&\omit&
&\omit&
\cr
\noalign{\hrule}
height3pt
&\omit&
&\omit&
&\omit&
\cr
&\hfil $\ZZ+{1\over 2}$&
&\hfil $2\ZZ$&
&\hfil $(C,S)$, $(S,C)$ &
\cr
height3pt
&\omit&
&\omit&
&\omit&
\cr
\noalign{\hrule}
height3pt
&\omit&
&\omit&
&\omit&
\cr
&\hfil $\ZZ+{1\over 2}$&
&\hfil $2\ZZ+1$&
&\hfil $(0,V),(V,0)$ &
\cr
height3pt
&\omit&
&\omit&
&\omit&
\cr
\noalign{\hrule}
height3pt
&\omit&
&\omit&
&\omit&
\cr
&\hfil $\ZZ$&
&\hfil $2\ZZ+1$&
&\hfil $(S,S)$, $(C,C)$ &
\cr
height3pt
&\omit&
&\omit&
&\omit&
\cr
}\hrule height 1.1pt
}
$$
}
\centerline{\sl Table 5: Covariant lattice representation of the
thermal $Spin(24)$
theory.  }

\noindent
This spectrum agrees precisely with the one given already in \twert.

The spectrum of the theory compactified with $(-)^{f_L}$ twist can be
obtained in an entirely
analogous manner, by including a Wilson-line $R\vec{A} = (\half^{12})
\in S$ of $Spin(24)$.
The untwisted
sector (even winding) is given by the decomposition of the covariant
lattice
$0\oplus C=(0,0)\oplus(V,V)\oplus (C,S)\oplus (S,C)$ while the twisted
sector (odd winding)
is the shifted lattice $(V,C)\oplus (0,S)\oplus (C,0)\oplus (S,V)$. The
$\vec{k_L}\cdot \vec{A}R$ is
half-integral for $\vec{k}_L\in V\oplus C$, so these are the conjugacy
classes that have
half-integral
momentum in the untwisted sectors. Since $\vec{A}^2R^2=3$ the changes
of modings
are the opposite in the twisted sector, {\it i.e.} the momentum is
half-integral for
$\vec{k}_L\in 0\oplus S$. This gives the spectrum
\bigskip
\vbox{
$$\vbox{\offinterlineskip
\hrule height 1.1pt
\halign{&\vrule width 1.1pt#
&\strut\quad#\hfil\quad&
\vrule width 1.1pt#
&\strut\quad#\hfil\quad&
\vrule width 1.1pt#
&\strut\quad#\hfil\quad&
\vrule width 1.1pt#\cr
height3pt
&\omit&
&\omit&
&\omit&
\cr
&\hfil $\tilde{n}$&
&\hfil $w$&
&\hfil $Spin(8,24)$&
\cr
height3pt
&\omit&
&\omit&
&\omit&
\cr
\noalign{\hrule height 1.1pt}
height3pt
&\omit&
&\omit&
&\omit&
\cr
&\hfil $\ZZ$&
&\hfil $2\ZZ$&
&\hfil $(0,0)$, $(C,S)$&
\cr
height3pt
&\omit&
&\omit&
&\omit&
\cr
\noalign{\hrule}
height3pt
&\omit&
&\omit&
&\omit&
\cr
&\hfil $\ZZ+\half$&
&\hfil $2\ZZ$&
&\hfil $(V,V)$, $(S,C)$ &
\cr
height3pt
&\omit&
&\omit&
&\omit&
\cr
\noalign{\hrule}
height3pt
&\omit&
&\omit&
&\omit&
\cr
&\hfil $\ZZ$&
&\hfil $2\ZZ+1$&
&\hfil $(V,C)$, $(S,V)$ &
\cr
height3pt
&\omit&
&\omit&
&\omit&
\cr
\noalign{\hrule}
height3pt
&\omit&
&\omit&
&\omit&
\cr
&\hfil $\ZZ+\half$&
&\hfil $2\ZZ+1$&
&\hfil $(C,0)$, $(0,S)$ &
\cr
height3pt
&\omit&
&\omit&
&\omit&
\cr
}\hrule height 1.1pt
}
$$
}
\centerline{\sl Table 6: Covariant lattice representation of the
$Spin(24)$ theory with $(-)^{f_L}$ twist.}

\noindent
which is precisely the spectrum given already in \newtwist.

The lattice implementation of the $(-)^{F_L}$ and $(-)^{f_L}$ twists
simply recasts the
discussion of the center of $Spin(4n)$ (section 2.3) in terms of Wilson
lines. The
identification is that ${\cal Z}_1\sim (\vec{A}R\in V)$ and ${\cal
Z}_2\sim (\vec{A}R\in S)$.
The advantage of the lattice technology is that it automates the
procedure; and this
is helpful when considering Wilson lines that are not in the center of
the gauge groups.
As an example of this, let us show that the $Spin(24)$ and the
$Spin(8)\times E_8$ theories are
related by T-duality $R\to 1/R$ after suitable Wilson lines are turned
on. The computation is
similar to the standard comparison between the $Spin(32)/\ZZ_2$ and
$E_8\times E_8$
heterotic theories in ten dimensions \GinspargBX. The computation (and
the relation to
10 dimensions) is clearest if we start from the covariant lattice
$0\oplus S$ \foot{As noted
in a previous footnote, this is related to the convention used in the
rest of the paper
through conjugaction by ${\cal C}$ defined in \Cdef.}.
The strategy is to decompose the covariant lattice $Spin(8,24)$ into
$Spin(8,8)\times Spin(16)$ and then add Wilson lines.

The covariant lattice of the $Spin(24)$ theory with conjugacy classes
$0\oplus S$  decomposes
to $\vec{w}_L\in(0,0)\oplus (V,V) \oplus (S,S) \oplus (C,C)$
under the $Spin(8,8)\times Spin(16)$ subgroup. The Wilson line
$R\vec{A} = (0^4;{1\over 2}^8)$ is in $S$ of $Spin(16)$ and leaves the
$Spin(8,8)$ intact.
The shifted lattice-vector $\vec{k}_L= \vec{w}_L - wR\vec{A}$ has the
spectrum
\eqn\klspec{
\eqalign{
& (0,0)\oplus(V,V) \oplus (S,S) \oplus (C,C)~~~~;~w\in 2\ZZ \cr
& (0,S)\oplus(S,0) \oplus (V,C) \oplus (C,V)~~~~;~w\in 2\ZZ+1
}}
The lattice vectors in the $X$-direction take the form
$p_{R,L} = {{\tilde n}\over R}\pm {wR\over 2}$ where
\eqn\ntildedef{
\tilde{n} = n - \vec{w}_L\cdot \vec{A} R + {w\over 2} \vec{A}^2 R^2
=  n - w- \vec{k}_L\cdot \vec{A} R
}
since $\vec{A}^2 R^2=2$.
The $\vec{k}_L\cdot \vec{A}R $ is integer (half-integer) when the
$Spin(16)$ part of
$\vec{k}_L\in Spin(8)\times Spin(16)$
is in $0\oplus S$ ($C\oplus V$). The complete spectrum of the $Spin(24)$
theory is then specified as
\bigskip
\vbox{
$$\vbox{\offinterlineskip
\hrule height 1.1pt
\halign{&\vrule width 1.1pt#
&\strut\quad#\hfil\quad&
\vrule width 1.1pt#
&\strut\quad#\hfil\quad&
\vrule width 1.1pt#
&\strut\quad#\hfil\quad&
\vrule width 1.1pt#\cr
height3pt
&\omit&
&\omit&
&\omit&
\cr
&\hfil $\tilde{n}$&
&\hfil $w$&
&\hfil $Spin(8,8)\times Spin(16)$ &
\cr
height3pt
&\omit&
&\omit&
&\omit&
\cr
\noalign{\hrule height 1.1pt}
height3pt
&\omit&
&\omit&
&\omit&
\cr
&\hfil $\ZZ$&
&\hfil $2\ZZ$&
&\hfil $(0,0)$, $(S,S)$ &
\cr
height3pt
&\omit&
&\omit&
&\omit&
\cr
\noalign{\hrule}
height3pt
&\omit&
&\omit&
&\omit&
\cr
&\hfil $\ZZ+{1\over 2}$&
&\hfil $2\ZZ$&
&\hfil $(V,V)$, $(C,C)$ &
\cr
height3pt
&\omit&
&\omit&
&\omit&
\cr
\noalign{\hrule}
height3pt
&\omit&
&\omit&
&\omit&
\cr
&\hfil $\ZZ$&
&\hfil $2\ZZ+1$&
&\hfil $(0,S)$, $(S,0)$ &
\cr
height3pt
&\omit&
&\omit&
&\omit&
\cr
\noalign{\hrule}
height3pt
&\omit&
&\omit&
&\omit&
\cr
&\hfil $\ZZ+{1\over 2}$&
&\hfil $2\ZZ+1$&
&\hfil $(C,V)$, $(V,C)$ &
\cr
height3pt
&\omit&
&\omit&
&\omit&
\cr
}\hrule height 1.1pt
}
$$
}
\centerline{\sl Table 7: Spectrum of the $Spin(24)$ theory with Wilson
line
$R\vec{A}=(0^4;{1\over 2}^8)$.}

Next, we consider the $Spin(8)\times E_8$ theory. Then the covariant
lattice decomposes to
$(0,0)\oplus (0,S) \oplus (S,0) \oplus (S,S)$ under the
$Spin(8,8)\times Spin(16)$
subgroup. The Wilson line $R\vec{A} = (1,0^3;1,0^7)$ belongs to the
$(V,V)$ conjugacy
class of the full $Spin(8,8)\times Spin(16)$. It shifts the
lattice-vector
$\vec{k}_L= \vec{w}_L - wR\vec{A}$ has the spectrum
\eqn\klspec{
\eqalign{
& (0,0)\oplus(0,S) \oplus (S,0) \oplus (S,S)~~~~~~~;~w\in 2\ZZ \cr
& (V,V)\oplus(C,C) \oplus (V,C) \oplus (C,V)~~~;~w\in 2\ZZ+1
}}
In this case $\vec{k}_L\cdot \vec{A}R$ is integer (half-integer) for
the diagonal (off-diagonal)
conjugacy classes. This shifts the allowed values of $\tilde{n}$
(defined in \ntildedef)
so that the spectrum becomes
\bigskip
\vbox{
$$\vbox{\offinterlineskip
\hrule height 1.1pt
\halign{&\vrule width 1.1pt#
&\strut\quad#\hfil\quad&
\vrule width 1.1pt#
&\strut\quad#\hfil\quad&
\vrule width 1.1pt#
&\strut\quad#\hfil\quad&
\vrule width 1.1pt#\cr
height3pt
&\omit&
&\omit&
&\omit&
\cr
&\hfil $\tilde{n}$&
&\hfil $w$&
&\hfil $Spin(8,8)\times Spin(16)$ &
\cr
height3pt
&\omit&
&\omit&
&\omit&
\cr
\noalign{\hrule height 1.1pt}
height3pt
&\omit&
&\omit&
&\omit&
\cr
&\hfil $\ZZ$&
&\hfil $2\ZZ$&
&\hfil $(0,0)$, $(S,S)$ &
\cr
height3pt
&\omit&
&\omit&
&\omit&
\cr
\noalign{\hrule}
height3pt
&\omit&
&\omit&
&\omit&
\cr
&\hfil $\ZZ$&
&\hfil $2\ZZ+1$&
&\hfil $(V,V)$, $(C,C)$ &
\cr
height3pt
&\omit&
&\omit&
&\omit&
\cr
\noalign{\hrule}
height3pt
&\omit&
&\omit&
&\omit&
\cr
&\hfil $\ZZ+{1\over 2}$&
&\hfil $2\ZZ$&
&\hfil $(0,S)$, $(S,0)$ &
\cr
height3pt
&\omit&
&\omit&
&\omit&
\cr
\noalign{\hrule}
height3pt
&\omit&
&\omit&
&\omit&
\cr
&\hfil $\ZZ+{1\over 2}$&
&\hfil $2\ZZ+1$&
&\hfil $(C,V)$, $(V,C)$ &
\cr
height3pt
&\omit&
&\omit&
&\omit&
\cr
}\hrule height 1.1pt
}
$$
}
\centerline{\sl Table 8: Spectrum of the $Spin(8)\times E_8$ theory
with Wilson line
$R\vec{A}=(1,0^3;1,0^7)$.}

The duality between the two heterotic theories can now be established
by comparing table
7 and table 8; they agree after taking $R\to 1/R$ and
$\tilde{n}\leftrightarrow w/2$ ({\it i.e.}
$n^\prime = w/2$ and $w^\prime = 2{\tilde n}$). This concludes the
explicit verification that
the two theories are on the same moduli space.

\listrefs

\end

The $Spin(24)$ theory is more confusing. Let us examine the
decompactification
limits in detail. As $R\to\infty$ we keep $w=0$ (and introduce
continuum notation for
the momenta)
\eqn\newtwi{\eqalign{
&T_V = e^{-\varphi} \bar{\cal O}_V(\half) V_{p} \cr
&\Psi_S = e^{-\varphi/2 + iH/2} \bar{\cal O}_S (\half) V_{p}~,
\quad\quad  p_R\geq 0~~\cr
&\Psi_C =e^{-\varphi/2 - iH/2} \bar{\cal O}_C (\half) V_{p}~,\quad\quad
p_R\leq 0
}}
while at $R\to 0$ we keep $p=0$ and rename $w\to p'$
\eqn\newtwis{\eqalign{
& T_C = e^{-\varphi} \bar{\cal O}_C (\half) V_{p'} \cr
&\Psi_S = e^{-\varphi/2 + iH/2} \bar{\cal O}_S (\half) V_{p'}~,
\quad\quad p_R\geq 0~~\cr
&\Psi_V =e^{-\varphi/2 - iH/2} \bar{\cal O}_V (\half)V_{p'}~,\quad\quad
  p_R\leq 0
}}
Formally, the theories are related by the interchange of the $V$ and
$C$ conjugacy
classes. However, in the $Spin(24)$ theory, only the $V$ conjugacy class
contains an operator with ${\bar\Delta}=1/2$. Thus, as $R\to\infty$, we
recover the $24$
propagating tachyons but, as $R\to 0$, there are $24$ fermions, all
propagating
to the left. Such a theory is anomalous and almost certainly
inconsistent.


The proliferation of string theories
in two dimensions is reminiscent
of the $5$ critical string theories  in $10$ dimensions which, along
with the elusive
$11$ dimensional M-theory, form the corners of the famous duality web,
encoding
the nonperturbative dualities between all string theories. It is now
clear that an
analogous duality web exists for two dimensional theories. Due to the
simplified kinematics,
and because non-perturbative definitions of the theories
exist, it should be possible to analyze the space of two-dimensional
theories in great
detail.

The full landscape of string theories is usefully thought of as the
space of all theories,
including non-critical ones (see {\it e.g.} \refs{\MaloneyRR,\AdamsRB});
so two-dimensional theories
are not just toy models, they are legitimate components of the full
landscape.
Indeed, it has been observed that closed string tachyon condensation,
generated by adding a
relevant operator to a critical string theory, typically has
non-critical string
theory as its end-point \HarveyNA. In this sense the two-dimensional
theories
form a particularly important part of the full landscape: they are the
generic ground states. It is clearly interesting to develop a more
complete picture of the space of two-dimensional theories.